\documentclass{article}
\usepackage{amsmath, amsthm, amssymb,latexsym,enumerate}
\usepackage{datetime}

\newcommand{\be}{\begin{equation}}
\newcommand{\ee}{\end{equation}}
\def\bea{\begin{eqnarray}}
\def\eea{\end{eqnarray}}
\def\bean{\begin{eqnarray*}}
\def\eean{\end{eqnarray*}}

\newcommand{\lab}{\boldsymbol\lambda}

\newcommand{\ov}{\overline}

\newcommand{\bcU}{\boldsymbol{\mc U}}

\def\sla{\raise.15ex\hbox{$/$}\kern-.57em}

\newcommand{\nn}{\nonumber}

\newcommand{\SM}{Standard Model}

\newcommand{\ba}{\begin{array}}
\newcommand{\ea}{\end{array}}
\newcommand{\mc}{\mathcal}

\begin{document}

%\begin{titlepage}
%\begin{flushright} %
%{\jobname}
%\framebox{~~ \ampmtime~--~ \today { }}
%\end{flushright}
%\vskip 2cm
%\begin{center}
%{\bf \Large THE FIVE INSTRUCTIONS}
%\end{center}
%\vskip 1cm

%\begin{center}{
%{\bf  Pierre Ramond${}^{a}_{}$}
%\vskip .9cm
%{\em Institute for Fundamental Theory,\\
%Department of Physics, University of Florida\\Gainesville FL 32611, USA}
%\vskip .2cm}\end{center}

%\vskip .9cm
%\begin{abstract}
%\noindent  
%Five Lectures on the Standard Model
%\end{abstract}

%\vskip 1.4cm

%\noindent Keywords: Superspace; Light-cone; Superconformal Theories; Chern-Simons Theories.

%\vspace{1cm}
%\vfill \vskip 5mm \hrule width 5.cm \vskip 2mm {\small
%\noindent $^a~$ramond@phys.ufl.edu
%}
%\end{titlepage}

%%%%%%%%%%%%%%%%%%%%%%%%
\centerline{\LARGE The Five Instructions}{
 %\vskip .3cm\centerline{\Large (TASI 2011)}
\begin{center}{
{\bf  Pierre Ramond}
\vskip .9cm
{\em Institute for Fundamental Theory,\\
Department of Physics, University of Florida\\Gainesville FL 32611, USA}
\vskip .2cm}\end{center}\vskip .5cm

%\centerline{Pierre Ramond}
%\vskip .5cm

\vskip .9cm
\begin{abstract}
\noindent  
Five elementary lectures delivered at TASI 2011on the Standard Model, its extensions to neutrino masses, flavor symmetries, and Grand-Unification.
\end{abstract}

%\vskip 1.4cm

%\noindent Keywords: Superspace; Light-cone; Superconformal Theories; Chern-Simons Theories.

%\vspace{1cm}
%\vfill \vskip 5mm \hrule width 5.cm \vskip 2mm {\small
%\noindent $^a~$ramond@phys.ufl.edu}

\section{Introduction}
In the once fast moving field of Particle Physics, the basic ingredients of the \SM ~were in place well before any of you were born! In 1976 (any of you 35 years old?), an epochal experiment (Prescott et al) provided the definitive evidence that all neutral current effects stemmed from one parameter, the Weinberg angle. Since then, experiments have confirmed the validity of the \SM~ without challenging its fundamental structure: it is a relativistic {\em local} quantum field theory in four space-time dimensions. 

Experiments have found all but one of its predicted particles:  in 1983, the $W^\pm$ and $Z$ gauge bosons are discovered at CERN; the $Z$ width suggests {\em at most} three ``normal" neutrinos. In 1995 the top quark is discovered, and the $\nu_\tau$ neutrino in 2000, both at FermiLab. The Higgs particle, a spinless particle predicted by the \SM~ remains beyond the grasp of experimentalists. 
\vskip .3cm
\centerline{\large There is more in Nature than the \SM:}
\vskip .2cm
\noindent - The first evidence is Dark Matter in the Universe, which suggests (a) new particle(s), stable enough to be fourteen billion years old, {\em not} predicted by the \SM.
\vskip .2cm
\noindent - The second chink in the armor of the \SM~appears in 1998, when the SuperKamiokande collaboration presents definitive evidence for neutrino oscillations, implying at least two massive neutrinos which also require new particle(s) {\em not} predicted by the \SM. 
\vskip .3cm
 
The LHC era begins in 2008 with a (mini) bang, not the one they were looking for! In record time, the LHC has experimentally rediscovered (most of) the \SM, but has not yet detected the Higgs, nor any particles indicating physics beyond the \SM. 

Like all successful theories, the \SM~offers new hints and poses new problems. Massive neutrinos {\em require} an extension of the \SM.  
The mystery of three families of elementary particles and their disparate masses and mixings, the {\em flavor problem} demands resolution.
The quantum numbers patterns offer the most spectacular hints, together with asymptotic freedom,  of a unified picture of quarks and leptons,  
at a scale close to the Planck length. Finally, although the \SM~ appears to be a mostly perturbative quantum field theory, it includes a relatively light spinless particle. This apparent theoretical curiosity (it is a puzzle if the \SM~is expanded to include new physics at much larger scales) has led to endless speculations, ranging from supersymmetry, to technicolor, to extra dimensions, and finally to  ``whatever theories" at  $TeV$ scales. 

The LHC discoveries will be cast in terms of the \SM: yesterday's theory is today's background. To prepare you for the new discoveries, it is necessary to have a sound understanding of the \SM, which is the purpose of these five instructions:  
 
\vskip .3cm
\noindent - First Instruction:  All About Relativistic Local  Field Theories
\vskip .2cm
\noindent - Second Instruction: The Standard Model
\vskip .2cm
\noindent - Third Instruction: The \SM~ with Massive Neutrinos
\vskip .2cm
\noindent - Fourth Instruction: Why Three Families?
\vskip .2cm
\noindent - Fifth Instruction: Grand Unification of the \SM
\vskip .3cm
Gravity is not included in the following discussion, since it is consistent with General Relativity, albeit with an inexplicably small ({\em eV} scale) cosmological constant.

There will be no references in the text. At the end of these instructions a (non-exhaustive) list of books I have found useful in the study of these topics.

\section{All About Relativistic Local Field Theory}
The \SM, which describes the Strong, Weak and Electromagnetic Interactions, appears to be a relativistic, renormalizable local quantum field theory, augmented by three different kinds of gauge symmetries. Its basic ingredients are  fields, local in space-time, transforming as representations of the Lorentz group.
\subsection{Lorentz Symmetry of Local Fields}
Elementary particles are described by local fields with definite transformation properties under the recognized symmetries of Nature. Those begin with the Lorentz symmetries of special relativity. Augmented by  space-time translations, they form the ten-parameters Poincar\'e group.  Its representations yield the space-time quantum numbers of particles, mass, momentum and spin. 

Local fields do not shift under space-time translations, so we need only consider the Lorentz transformations. They are generated by three rotations $J^{}_i$ and three space vector boosts $K_i^{}$, which satisfy the commutation relations

$$
[\,J^{}_i\,,\,J^{}_j\,]~=~i\epsilon^{}_{ijk}\,J^{}_k\ ,\qquad [\,J^{}_i\,,\,K^{}_j\,]~=~i\epsilon^{}_{ijk}\,K^{}_k\ ,
$$
as well as 

$$
[\,K^{}_i\,,\,K^{}_j\,]~=-i\epsilon^{}_{ijk}\,J^{}_k\ .
$$
with $i,j,k=1,2,3$. We can express these in terms of two sets of generators, 

$$
L^{}_i~=~\frac{1}{\sqrt{2}}(J^{}_i+iK^{}_i)\ ,\qquad R^{}_i~=~\frac{1}{\sqrt{2}}(J^{}_i-iK^{}_i)\, $$
each satisfying the commutation relations of $SU(2)$,

$$ 
[\,L^{}_i\,,\,L^{}_j\,]~=~i\epsilon^{}_{ijk}\,L^{}_k\ ,\qquad [\,R^{}_i\,,\,R^{}_j\,]~=~i\epsilon^{}_{ijk}\,R^{}_k\ ,
$$
and commuting with one another 

$$ [\,L^{}_i\,,\,R^{}_j\,]~=~0\ .$$
There is one ``small" complication, however, neither $L^{}_i$ nor $R^{}_i$ are hermitean. Instead, they are related by conjugation since 
$L^\dagger_i~=~R^{}_i \ ,$
as well as by parity since  the angular momentum generators  are invariant under parity while the boosts change sign. 

This enables us to catalog all representations of the Lorentz group using the representation theory of $SU(2)$ which you know since third grade. We label the representations $(\,spin\,,\,spin'\,)$ under $SU(2)_L\times SU(2)_R$:

\bean
{\rm Scalar} &&(\,0\,,\,0\,)~\rightarrow~ \varphi(x)\\
{\rm Left-handed ~Spinor}&& (\,\frac{1}{2}\,,\,0\,)~\rightarrow ~\psi^{}_\alpha\ ,\quad \alpha=1,2\\
{\rm Right-handed ~Spinor}&& (\,0\,,\,\frac{1}{2}\,)~\rightarrow ~\psi_{}^{\dot\alpha}\ ,\quad \dot\alpha=1,2\\
{\rm Gauge~ Potential}&& (\,\frac{1}{2}\,,\,\frac{1}{2}\,)~\rightarrow ~A^{}_\mu\\
{\rm Self-Dual~ Field~ Strength }&& (\,1\,,\,0\,)\ ,~(\,0\,,\,1\,)~\rightarrow ~F^{}_{\mu\nu}=i\widetilde F^{}_{\mu\nu}\ ,~F^{}_{\mu\nu}=-i\widetilde F^{}_{\mu\nu}
\eean
and so on. Because of charge conjugation, a right-handed spinor can be written as a left-handed spinor. Under a Lorentz transformation, left and right spinors transform as

$$\psi^{}_{L(R)}~\rightarrow~\exp{ \Big[\frac{\sigma_i}{2}(\omega_i\mp i\nu_i)\Big]}\,\psi^{}_{L(R)}\ ,
$$
where $\omega_i$ and $\nu_i$ are the rotation and boosts parameters, respectively. The reader can check that the combination $\sigma_2\psi^*_R\sim \epsilon^{\alpha\beta}_{}\big(\psi^{*}_R\big)^{}_\beta$ transforms as a {\em left-handed} spinor. The discrete operation $P$ and $C$ are as follows

$$P:~~~\psi^{}_{L(R)}~\rightarrow~\psi^{}_{R(L)}\ ,~~C:~~\psi^{}_{L(R)}~\rightarrow~\sigma^{}_2\psi^*_{R(L)}\ ,$$
so that the combined $CP$ is sustained on one spinor,

$$
CP:~~\psi^{}_{L(R)}~\rightarrow~\sigma^{}_2\psi^*_{L(R)}\ .$$
Henceforth, we exclusively deal with spinors that transform left-handedly, which entails writing the right-handed spinor in terms of its conjugate,

$$\psi^{}_L~\rightarrow~ \psi\, \qquad \psi^{}_R ~\rightarrow~ \ov\psi\ .$$
It may be confusing but $\ov\psi$ is {\bf not} the conjugate of $\psi$; it is the conjugate of its right-handed partner. For example, the electron is written as {\em two} left-handed fields, $e(x)$ and $\ov e(x)$.
 
The translation-invariant derivative operator $\partial_\mu=(\frac{\partial}{\partial t},\frac{\partial}{\partial \vec x})$  allows for simple Lorentz-invariant combinations 

$$
\partial^{}_\mu\varphi\partial^\mu_{}\varphi\ ,\quad \psi^\dagger\sigma^\mu_{}\partial^{}_\mu\psi\ ,\quad \partial^{}_\mu A^{}_\nu\partial^\mu_{}A^\nu_{}\ ,\cdots
$$
where $\sigma^\mu=(1,\vec\sigma)$. Their real parts serve as the canonical kinetic terms for the scalar, spinor and gauge potentials. 

%%%%%%%%%%%%%%%%%%%%%%%%%%
\subsection{Gauge Symmetries of Local Fields}
The spinor kinetic term has an additional {\em global} invariance

$$\psi~\rightarrow~e^{i\lambda}_{}\psi\ ,
$$
as long as $\lambda$ does not depend on $x$. If the rest of the Lagrangian respects this symmetry, Noether's theorem tells us that it corresponds to a conservation law, in this case the  ``$\psi$-number''.

In order to generalize this symmetry to the local case, with $\lambda(x)$ as parameter, we need to generalize the derivative operator to a {\em covariant derivative} $\mc D_\mu$ in such a way that

$$
\psi~\rightarrow~e^{i\lambda(x)}_{}\psi\ ,\quad {\mc D}^{}_\mu\psi~\rightarrow~e^{i\lambda(x)}_{}{\mc D}^{'}_\mu\psi\ .
$$
As an operator equation,

$${\mc D}^{}_\mu~\rightarrow~{\mc D}^{'}_\mu~=~e^{i\lambda(x)}_{}{\mc D}^{}_\mu e^{-i\lambda(x)}_{}\ .
$$
This is solved by introducing a gauge potential $A_\mu(x)$, and setting

$${\mc D}^{}_\mu~\equiv~\partial^{}_\mu+iA^{}_\mu\ , \qquad A^{}_\mu(x)~\rightarrow~A^{'}_\mu(x)~=~A^{}_\mu(x)-\partial^{}_\mu\lambda(x)\ .
$$
This is like the gauge transformation of electrodynamics, of course. To give legs to the gauge potential, we must write a kinetic term that is both gauge and Lorentz invariant. We define the field strength

$$F^{}_{\mu\nu}~\equiv-i[\,{\mc D}^{}_\mu\,,\,{\mc D}^{}_\nu\,]\ ,$$
which is gauge- but not Lorentz-invariant. This is easily remedied by forming the combinations

$$
F^{}_{\mu\nu}F^{\mu\nu}_{}\ ,\qquad F^{}_{\mu\nu}\widetilde F^{\mu\nu}_{}\ .$$
The former is the desired kinetic term, while the latter is a total derivative which does not contribute to the equations of motion.

The Parity conserving version of this theory is QED (with massless electron),

$${\mc L}~=-\frac{1}{4e^2}F^{}_{\mu\nu}F^{\mu\nu}_{}+i\psi^\dagger\sigma^\mu_{}{\mc D}^{}_\mu\psi+i\ov\psi^\dagger\sigma^\mu_{}\ov{\mc D}^{}_\mu\ov\psi\ ,
$$
where 

$$\ov{\mc D}^{}_\mu~=~\partial^{}_\mu-iA^{}_\mu\ ,\quad {\rm since}~~~\ov\psi~\rightarrow~e^{-i\lambda}_{}\ov\psi\ .$$
You are probably more familiar with the Lagrange density written in terms of the four-component Dirac spinor and the $(4\times 4)$ Dirac matrices as  

$$i\ov\Psi_D\gamma_{}^\mu{\mc D}_\mu\Psi_D\ ,\qquad \Psi_D~=~\begin{pmatrix}\psi_L\\ \psi_R\end{pmatrix}\ .$$
\vskip .3cm
\noindent Yang \& Mills generalized this construction to the case of several spinor fields. The kinetic term for $N$ canonical spinor fields $\psi_a,\, a=1,2,\dots,N$, is simply

$$
i\sum_{a=1}^N \psi^{a\dagger}_{}\sigma^\mu_{}\partial^{}_\mu\psi^{}_a~=~i\Psi^\dagger_{}\sigma^\mu_{}\partial^{}_\mu\Psi\ .$$
It is invariant under the global phase transformation as before, but  has a much larger global invariance, namely

$$
\Psi~\rightarrow~{\bcU}\Psi\ ,$$ 
where $\bcU$ is an $(N\times N)$ unitary matrix, ${\bcU}^\dagger_{}{\bcU}~=~1$, with $(N^2-1)$ parameters. To see this, write it as the exponential of $i$ times a Hermitean matrix

$${\bcU}~=~e^{i\bf H}_{}\ ,\quad {\bcU}^\dagger_{}~=~e^{-i\bf H}_{}={\bcU}^{-1}_{}\ . $$
$\bf H$ has $(N-1)$ real diagonal elements (taking out the trace which is the overall phase), and $N(N-1)/2$ complex off-diagonal entries, totaling $(N^2-1)$ parameters. We can expand it in terms of $(N^2-1)$ hermitean matrices,

$$
{\bf H}~=~\frac{1}{2}\sum_{A=1}^{N^2-1}\omega^{}_A\,\lab^{}_A\ ,
$$
where the $\lab_A$ are the $(N^2-1)$ traceless hermitean matrices. They satisfy the Lie algebra, 

$$[\,\lab_A\,,\,\lab_B\,]~=~if^{}_{ABC}\lab^{}_C\ ,\qquad {\rm Tr}(\lab_A\lab_B)~=~\frac{1}{2}\delta^{}_{AB}\ ,$$
where $f^{}_{ABC}$ are the structure functions of $SU(N)$. For $N=2$ they reduce to the Pauli matrices; for $N=3$, they are the Gell-Mann matrices, and so-on.

The upgrade of these global transformations to space-time dependence requires a covariant derivative, but this time it is a $(N\times N)$ matrix operator, with the properties

$$
{\boldsymbol{\mc D}}^{}_\mu~\rightarrow~{\boldsymbol{\mc D}}^{'}_\mu~=~{\bcU}\,{\boldsymbol{\mc D}}^{}_\mu\,{\bcU}^\dagger_{}\ .
$$ 
Set 

$$
{\boldsymbol{\mc D}}^{}_\mu~=~\partial^{}_\mu{\bf 1}+i{\bf A}^{}_\mu\ ,$$
where $\bf A_\mu$ is now a hermitian matrix with $(N^2-1)$ gauge potentials, 

$$
{\bf A}^{}_\mu~=~\frac{1}{2}\sum_{B=1}^{N^2-1}\,A^B_\mu\lab_B^{}\ ,$$
which behave under gauge transformations as,

$$
{\bf A}^{}_\mu~\rightarrow~{\bf A}^{'}_\mu~=-i\,{\bcU}\,\partial^{}_\mu{\bcU}^\dagger_{}-i\,{\bcU}\,{\bf A}^{}_\mu\,{\bcU}^\dagger_{}\ ,$$
which generalizes the QED gauge transformation to the non-Ablian case. To obtain the field strengths, we form the commutator

$$
{\bf F}^{}_{\mu\nu}~=~-i[\,{\boldsymbol{\mc D}}^{}_\mu\,,\,{\boldsymbol{\mc D}}^{}_\nu\,]\ ,$$
or in terms of components

$$
{F}^{B}_{\mu\nu}~=~\partial^{}_\mu A^B_\nu-\partial^{}_\nu A^B_\mu-f^{BCD}_{}\,A^C_\mu A^D_\nu\ .$$
They are not gauge-invariant, but do transform nicely (covariantly)

$$
{\bf F}^{}_{\mu\nu}~\rightarrow~\bcU\,{\bf F}^{}_{\mu\nu}\bcU^\dagger_{}\ ,$$
making it easy to form the gauge- and Lorentz-invariant kinetic term

$$-\frac{1}{2g^2}{\rm Tr} \Big({\bf F}^{}_{\mu\nu}{\bf F}_{}^{\mu\nu}\Big)\ .
$$

Let us now specialize to $N=3$ and parity invariant interactions with one spinor transforming as triplet. The Lagrange density would be

$$
{\mc L}~=-\frac{1}{2g^2}{\rm Tr} \Big({\bf F}^{}_{\mu\nu}{\bf F}_{}^{\mu\nu}\Big)+i\Psi^\dagger_{}\sigma^\mu_{}{\boldsymbol{\mc D}}^{}_\mu\Psi+i\ov\Psi^\dagger_{}\sigma^\mu_{}\ov{\boldsymbol{\mc D}}^{}_\mu\ov\Psi\ ,$$
which is the QCD Lagrangian of eight gauge potentials (gluons) interacting with coupling $g$ with one quark. 

Finally, note that although we have constructed the covariant derivative from spinors transforming as the $N-$dimensional representation of $SU(N)$, it can be generalized to {\em any} representation of the algebra. Acting on fields which transform as the $D$-dimensional representation of $SU(N)$, the covariant derivative is a $(D\times D)$ matrix operator,
$$ 
{\boldsymbol{\mc D}}^{}_\mu~=~\partial^{}_\mu{\bf 1}+i\sum_{B=1}^{N^2-1}{A}^{B}_\mu{\bf T}^B\ ,$$
where ${\bf T}^B$ are the $(D\times D)$matrices which satisfy the $SU(N)$ Lie algebra. 

\subsection{Renormalizability and All That} 
A bosonic  field $\varphi(x)$, or fermionic field $\psi(x)$  is canonical if its kinetic term in the Action is of the form 

$$\int d^4x \partial_\mu\varphi(x)\partial^\mu\varphi(x)\ ,\qquad i\int d^4x \psi^\dagger(x)\sigma_\mu\partial^\mu\psi(x)\ .$$
Since the Action is dimensionless, the bosonic and fermionic fields have (mass) dimensions 

$$[\varphi]~=~1\ ,\qquad [\psi]~=~\frac{3}{2}\ .$$

The Action contains all interactions (polynomials) in those fields, consistent with the invariances of the theory. They are most easily catalogued by their mass dimensions. In four dimensions, polynomials with mass dimension four are called {\em marginal}; those with dimensions less than four are called {\em relevant} (in the infrared where masses dominate), and finally those with dimensions greater than four are called {\em irrelevant} (in the infrared). This unfortunate notations comes from condensed matter where infrared is more relevant than ultraviolet!

To make dimensionless contributions to the Action, relevant interactions are multiplied by positive powers of mass,  irrelevant interactions by negative powers. Naively, in the limit of large masses, only relevant interactions survive and vice-versa. This heuristic feature survives quantum effects. Marginal interactions appear multiplied by dimensionless parameters, which are the coupling ``constants" of the theory.

The Action of a renormalizable field theory contains only relevant and marginal interactions. When such an Action (with external sources) is inserted in the Dirac-Feynman path integral, it yields the Schwinger Effective Action, a generalization of the classical Action. Besides the original relevant and marginal interactions, it contains an infinite number of ``irrelevant interactions with higher powers of derivatives and fields stemming from quantum effects. In a renormalizable theory, the ultraviolet divergences have been absorbed through a careful redefinition of the input fields and parameters. As a result, the coefficients in front of the irrelevant interactions in the effective Lagrangian are {\em calculable and finite}.

\subsubsection{QED}
It is instructive to see how this translates in the presentation of QED, the simplest and most familiar gauge theory. The basic ingredients are the electron field, the gauge field (potential) $A_\mu(x)$, a dimensionless coupling ``constant" and the electron mass. Since parity is conserved, there are two Weyl spinor fields, $\psi_L(x)$ and $\psi_R(x)$ which combine into one four-component Dirac field $\Psi(x)$, to represent the electron and positron.
At the classical level, besides the usual kinetic terms, its marginal and relevant interactions are, respectively,  

$$e\,\ov\Psi(x)\gamma_\mu\Psi(x)\, A^\mu_{}(x)\ ,\quad m^{}_e\ov\Psi(x)\Psi(x)\ ,$$ 
where $m_e$ is the electron mass. It is renormalizable since it contains no irrelevant interactions. The mass and coupling parameters are not constants as their values depend on the energy at which they are measured. They run with scale, and satisfy the one-loop renormalization group equations

$$\frac{d\,e}{dt}~=~\frac{e^3}{12\pi^2}\ ,\qquad \frac{d\,m^{}_e}{dt}~=-\frac{3e^2_{}}{8\pi^2}m^{}_e\ .$$
where $t$ is the logarithm of the energy scale. Hence the coupling gets larger and larger with the energy at which it is measured, leading to the Landau pole. {\em Caveat Emptor}: as $e$ increases, we quickly lose the ability to do perturbation theory; all we can say is ``when last seen, the coupling is getting bigger". 

Note that the mass parameter does not run if we set $m_e=0$. This is indicative that this point is stable, indicating the presence of a symmetry respected by quantum effects. 't Hooft calls $m_e=0$ a {\em natural value}: a natural value is one at which symmetry is larger (the symmetry is chiral symmetry-more on this later). 

Without elaborate calculations, we can anticipate which irrelevant interactions are generated by quantum effects. It is best to order them in terms of dimensions.
\vskip .2cm
\noindent {\em Dimension Five Interactions}. Gauge invariance requires  the gauge potential to appear in covariant derivatives only.  There are no Lorentz-invariant polynomials of dimension-five built out solely out of dimension-one covariant derivative. Fermions must be present  in even numbers; there are several types of dimension-three fermion bilinears 

$$
\ov\Psi\,\Psi\ ,~~\ov\Psi\,\gamma_5\Psi\ ,~~\ov\Psi\,\gamma_\mu\Psi\ ,~~\ov\Psi\,\gamma_{\mu\nu}\Psi\ ,~~\ov\Psi\,\gamma_\mu\gamma_5\Psi\ ,$$
some of which, paired with two covariant derivatives, can yield dimension-five Lorentz-invariant combinations 

$$
\ov\Psi\,{\mc D}_\mu{\mc D}^\mu\Psi\,\ ,\quad  \ov\Psi\,\gamma_{\mu\nu}\Psi\,F^{\mu\nu}_{}\ ,\quad \ov\Psi\,\gamma_{5}{\mc D}_\mu{\mc D}^\mu\Psi\,\ ,\quad \ov\Psi\,\gamma_{\mu\nu}\Psi\,\widetilde F^{\mu\nu}_{}\ .$$
The last two interactions {\em cannot} be generated in the effective Action because of they are odd under $P$ and $CP$, which QED respects! Otherwise the electron would have an intrinsic electric dipole moment. 

The first two interactions, allowed by all QED  symmetries, are generated by loop effects.  The most famous one describes the interaction between the electron's magnetic moment and the magnetic field. It is easy to see that the lowest order Feynman diagram which generates this interaction is proportional to $\alpha$, the dimensionless fine structure ``constant", leading to 

$$\Big(\frac{\bf 1}{\boldsymbol\pi}\Big)\frac{\alpha}{m_e}\ov\Psi\,\gamma_{\mu\nu}\Psi\,F^{\mu\nu}_{}\ .$$
The hard part is the calculation of the prefactor $1/\pi$. It is of course to Schwinger's  famous result for the electron Land\'e g-factor, $g-2=\frac{\alpha}{2\pi}$. This is the lowest order contribution, as there are similar terms in higher order in $\alpha$, and an infinite number of higher dimension terms with   more covariant derivatives. The other interaction yields chirality-breaking finite corrections to the electron mass, a momentum-dependent electron-photon vertex,  and a four-point  photon-electron interaction.

\vskip .2cm
\noindent {\em Dimension Six Interactions}. All  appear divided by the $m_e^2$. These can contain six covariant derivatives, two fermions and three covariant derivatives, or four fermions. 
There is only one possible interaction without fermions (up to integration by parts),

$$
{\mc D}^{}_\mu F^{\mu\nu}_{}{\mc D}^{}_\rho F^{\rho}_{~\nu}\ ,$$
which corrects the photon propagator. The three-field strength interaction is forbidden by Furry's theorem (no vertex with an odd number of photons).  There are several  new interactions with two fermions; we only note 

$$\ov\Psi\gamma_\mu^{}\Psi {\mc D}^{}_\nu F^{\mu\nu}_{}\ ,$$
which contributes to the form factor of the electron-photon vertex, and yields the charge radius of the electron.
%$$\ov\Psi\gamma^{}_\mu\Psi\partial^{}_\nu F^{\mu\nu}_{}\ ,$$

This leaves four-fermion interactions. After Fierzing,  and taking into account parity conservation, only three terms remain, 

$$
(\ov\Psi\,\Psi)^2_{}\ ,\quad(\ov\Psi\,\gamma_5\Psi)^2_{}\ ,\quad\ov\Psi\,\gamma_\mu\Psi\ov\Psi\,\gamma^\mu\Psi\ .$$ 
They are generated by one-loop ``box diagrams".  In QED, they correct the four-fermion interactions which already exist at tree level, and thus do not give rise to new types of interactions. This will change in the \SM~ where box diagrams give rise to new interactions that are not present at tree-level.

\vskip .2cm
\noindent {\em Dimension Eight Interactions}. QED generates brand new interactions at this level. They give rise to my favorite interaction, the scattering of light by light, which contains four field strengths. There are two such terms, with contribution in the effective Action

$$\frac{\alpha^2}{m^4_e}\Big(\frac{\bf 1}{\bf 90} (F^{}_{\mu\nu}F^{\mu\nu}_{})^2_{}+\frac{\bf 7}{\bf 270}(F^{}_{\mu\nu}\widetilde F^{\mu\nu}_{})^2_{}\Big)\ ,$$ 
where the numerical coefficients in bold are calculated from a one-fermion loop. When $m_e=0$,  infrared divergences spoil this analysis, and they have to be properly cut-off to carry out the same analysis. 

In conclusion, this method of listing all possible interactions dimension by dimension is a potent way to anticipate the effect of quantum corrections. Most, as we have seen, correct the tree-level interactions, but a few generate new interactions not seen at tree-level. 

%%%%%%%%%%%%%%%%%%%%%%%%%
\section{The Standard Model}
Now that we have assembled the pieces, we can write down the \SM~Lagrangian. It consists of three gauge theories based on the groups $SU(3)$ for color, $SU(2)$ for weak charge, and $U(1)$ for hypercharge, with three independent coupling constants $g_3$, $g_2$, and $g_1$, and three gauge potentials:  eight gluons, $G^A_\mu$, $A=1,2,\dots, 8$, three weak bosons, $W_\mu^a$, $a=1,2,3$, and one Hyperon $B^{}_\mu$, with field strengths

\bea
G^A_{\mu\nu}&=&\partial^{}_\mu G^A_\nu-\partial^{}_\nu G^A_\mu-f^{ABC}_{}G^B_\mu G^C_\nu\ ,\nn\\ 
W^a_{\mu\nu}&=&\partial^{}_\mu W^a_\nu-\partial^{}_\nu W^a_\mu-\epsilon^{abc}_{}W^b_\mu W^c_\nu\ ,\nn\\
B^{}_{\mu\nu}&=&\partial^{}_\mu B^{}_\nu-\partial^{}_\nu B^{}_\mu\ ,
\eea
where $f^{ABC}$ and $\epsilon^{abc}$ are the $SU(3)$ and $SU(2)$ structure functions. The first part of the \SM ~Lagrangian is then

$${\mc L}^{}_{YM}~=-\frac{1}{4g^2_3}G^A_{\mu\nu}G^{\mu\nu\,A}_{}-\frac{1}{4g^2_2}W^a_{\mu\nu}W^{\mu\nu\,a}_{}-\frac{1}{4g^2_1}B^{}_{\mu\nu}B^{\mu\nu}_{}\ .$$
Its matter content consists of  three families of spin one-half quarks and leptons (written as two component left-handed Weyl spinors) with identical gauge structures:

 \bean
 {\rm Lepton~doublet~}&&L_i^{}=\begin{pmatrix}\nu_i\\ {\rm e}_i\end{pmatrix}  \sim(\,{\bf 1}^c\,,\,{\bf 2}\,)^{}_{-1}\ ;\quad {\boldsymbol{\mc D}}^{}_\mu L^{}_i=(\partial^{}_\mu+i{\bf W}^{}_\mu-\frac{i}{2}B^{}_\mu)L_i\\
{\rm Lepton~singlet}\,~~&& \ov {\rm e}^{}_i\sim(\,{\bf 1}^c\,,\,{\bf 2}\,)^{}_{2}\ ; \hskip 2cm {\boldsymbol{\mc D}}^{}_\mu \ov {\rm e}^{}_i=(\partial^{}_\mu+iB^{}_\mu)\ov {\rm e}^{}_i\\
{\rm Quark~doublet}\,~~&& {\bf Q}_i^{}=\begin{pmatrix}{\bf u}_i\\ {\bf d}_i\end{pmatrix}\sim(\,{\bf 3}^c\,,\,{\bf 2}\,)^{}_{\frac{1}{3}}\ ;~ {\boldsymbol{\mc D}}^{}_\mu {\bf Q}^{}_i=(\partial^{}_\mu+i{\bf G}^{}_\mu+i{\bf W}^{}_\mu+\frac{i}{3}B^{}_\mu){\bf Q}_i\\
{\rm Quark~singlet}\,~~~&& \ov{\bf u}_i^{}\sim(\,{\bf 3}^c\,,\,{\bf 1}\,)^{}_{-\frac{4}{3}}\ ; \hskip 1.6cm {\boldsymbol{\mc D}}^{}_\mu \ov{\bf u}^{}_i=(\partial^{}_\mu-i{\bf G}^{*}_\mu-\frac{2i}{3}B^{}_\mu)\ov{\bf u}_i\\
{\rm Quark~singlet}~~~&& \ov{\bf d}_i^{}\sim(\,{\bf 3}^c\,,\,{\bf 1}\,)^{}_{\frac{2}{3}}\ ; \hskip 1.8cm {\boldsymbol{\mc D}}^{}_\mu \ov{\bf d}^{}_i=(\partial^{}_\mu-i{\bf G}^{*}_\mu+\frac{i}{3}B^{}_\mu)\ov{\bf d}_i\ ,
\eean
where $i=1,2,3$ runs over three families, and 

$$
{\bf G}^{}_\mu=G^A_\mu\frac{\lab^A}{2}\ ,~~ {\bf G}^{*}_\mu=G^A_\mu\frac{\lab^{*A}}{2}\ ,~~{\bf W}^{}_\mu=W^a_\mu\frac{\tau^a}{2}\ .$$
The values of the hypercharge $Y$have been adjusted so as to satisfy the Gell-Mann-Nishijima formula,
 
$$
Q~=~I^{}_3+\frac{Y}{2}\ .$$ 
Remarkably, these hypercharge values satisfy the anomaly conditions, that is the one fermion loop diagrams with three external gauge bosons of the form $U(1)\cdot U(1)\cdot U(1)$, $U(1)\cdot SU(2)\cdot SU(2)$, $U(1)\cdot SU(3)\cdot SU(3)$, all vanish. In addition the sum of all fermion hypercharges vanish as well, which can be viewed as the mixed gravitational anomaly! The \SM~is ready for the big time!

The gauge-fermion Lagrangian is the sum of all fermion kinetic terms with covariant derivatives.

$$
{\mc L}_D^{}~=i\sum_{i=1}^3\,\Big(L^\dagger_i\sigma^\mu_{}{\boldsymbol{\mc D}}^{}_\mu L^{}_i+\ov e^\dagger_i\sigma^\mu_{}{\boldsymbol{\mc D}}^{}_\mu \ov e^{}_i+ {\bf Q}^\dagger_i\sigma^\mu_{}{\boldsymbol{\mc D}}^{}_\mu {\bf Q}^{}_i+\ov{\bf u}^\dagger_i\sigma^\mu_{}{\boldsymbol{\mc D}}^{}_\mu \ov{\bf u}^{}_i+\ov{\bf d}^\dagger_i\sigma^\mu_{}{\boldsymbol{\mc D}}^{}_\mu \ov{\bf d}^{}_i\Big)\ .
$$
All gauge couplings respect chirality: left- and right-handed fermions evolve separately in the absence of masses, but in the real world, most fermions  and  some of the gauge bosons have masses. Masses can be introduced elegantly, without spoiling renormalizability, by adding a spin zero field 

$$
{\rm Higgs~ doublet}~~H=\begin{pmatrix} h_1^{}\\ h^{}_2\end{pmatrix}  \sim(\,{\bf 1}^c\,,\,{\bf 2}\,)^{}_{1}\ ;\quad {\boldsymbol{\mc D}}^{}_\mu H=(\partial^{}_\mu+i{\bf W}^{}_\mu+\frac{i}{2}B^{}_\mu)H\ .
$$
The Higgs-gauge Lagrangian is augmented by a potential

$$
{\mc L}^{}_H~=~({\boldsymbol{\mc D}}^{}_\mu H)^\dagger_{} {\boldsymbol{\mc D}}^{\mu}_{} H~+~m^2_{}H^\dagger H~-~\lambda(H^\dagger H)^2_{}
\ .$$
cleverly engineered pragmatically, to be renormalizable and yield a non vanishing value at minimum.  
For all the beauty of the gauge couplings, this is clearly the soft underbelly of the \SM. Nevertheless, it works nicely, at least at tree-level. Indeed minimization of the potential yields,

$$
 H_0^\dagger H^{}_0~=~\frac{m^2_{}}{2\lambda}\equiv \frac{v^2_{}}{2}\ ,~\rightarrow~ H^{}_0=\frac{1}{\sqrt{2}}\begin{pmatrix} 0\\ v\end{pmatrix}\ .
$$
In this configuration, the $SU(2)\times U(1)$ symmetry is {\em spontaneously} broken down to $U(1)$, as we can check with the Gell-Mann Nishijima formula. The Higgs field needs to be expanded away from its vacuum value in terms of fields which have no vacuum values, thereby recovering the usual perturbative expansion. To that effect, we introduce the very convenient Kibble parametrization,

$$
H~=~\frac{1}{\sqrt{2}}\bcU (x)\begin{pmatrix} 0\\ v+h(x)\end{pmatrix}\ ,
$$
where $\bcU (x)$ is the unitary matrix which contains the three Nambu-Goldstone modes generated by the spontaneous breaking. Remembering the magic properties of covariant derivatives, 

$$
{\boldsymbol{\mc D}}_{\mu}^{} H~=~{\boldsymbol{\mc D}}_{\mu}^{}\bcU (x)\begin{pmatrix} 0\\ v+h(x)\end{pmatrix}
~=~\frac{1}{\sqrt{2}}\bcU (x){\boldsymbol{\mc D}}_{\mu}^{'}\begin{pmatrix} 0\\ v+\rho(x)\end{pmatrix}\ ,
$$
where the prime on the covariant derivative refers to the gauge-transformed potentials. As a result, $\bcU (x)$ disappears from the Higgs Lagrangian, leaving just

$$
{\mc L}^{}_H~=~\frac{1}{2}\partial^{}_\mu h\partial^\mu_{}h-\frac{1}{2}M^{2}_h h^2_{}-M^{}_h\sqrt{\frac{\lambda}{2}}h^3_{}-\frac{\lambda}{4}h^4_{}+\frac{1}{8}(v+h)^2_{}
\Big( (B^{}_\mu-W^3_\mu)^2+|W^1_\mu+iW^2_\mu|^2_{}\Big)\ .
$$
It describes the Higgs boson, a scalar field $h$ of mass $M_h=v\sqrt{2\lambda}$ with cubic and quartic self-interactions. By rescaling the gauge fields ${\bf W}_\mu\rightarrow g_2{\bf W}_\mu$, ${B}_\mu\rightarrow g_1{B}_\mu$,  the gauge couplings to the Higgs  yield

$$
\frac{1}{2}M^2_Z Z^{}_\mu Z^\mu_{}+M^2_WW^+_\mu W^{-\,\mu}_{}+\frac{h}{v}(2+\frac{h}{v})\Big(\frac{1}{2}M^2_Z Z^{}_\mu Z^\mu_{}+M^2_WW^+_\mu W^{-\,\mu}_{}\Big)\ ,$$
where

$$
Z^{}_\mu=\cos\theta_W^{}W^3_\mu-\sin\theta^{}_WB^{}_\mu\ ,\qquad W^\mp_\mu~=~\frac{1}{\sqrt{2}}(W^1_\mu\pm iW^2_\mu)\ ,
$$
where $\theta_W$ is the Weinberg angle,   

$$\tan\theta^{}_W~=~\frac{g_1}{g_2}\ ,\quad M^{}_Z~=~\frac{v}{2}\sqrt{(g_1^2+g^2_2)}\ ,\quad M^{}_W~=~\cos\theta^{}_W\,M{}_Z\ .$$ 
The combination orthogonal to the $Z$-boson is of course to be identified with the massless photon

$$
A^{}_\mu=\sin\theta^{}_WW^3_\mu+  \cos\theta_W^{}B^{}_\mu\ .$$
This is the Higgs mechanism, invented by Brout and Englert: spontaneous breaking of gauged symmetries generate masses for the gauge bosons corresponding to the broken symmetries (they ``eat" the Nambu-Goldstone bosons which morph into the longitudinal modes of the massive gauge bosons). 
Note that the higgs boson couples to the gauge bosons according to their masses. 

The next step is to express the gauge fermion couplings to $A_\mu$, $Z_\mu$ and $W^\pm_\mu$. This is a  straightforward albeit tedious algebraic exercise. We just quote the results,

$${\mc L}^{\rm currents}~=~eA_{}^\mu \,J^{\rm em}_\mu ~+~g^{}_Z Z^{\mu}_{}\,J^{(Z)}_\mu~+~g^{}_W W^{+\,\mu}_{}\,J^-_\mu~+~g^{}_W W^{-\,\mu}_{}\,J^+_\mu\ ,$$
where the three current couplings are

$$e~=~\frac{g_1g_2}{\sqrt{g^2_1+g^2_2)}}\ ,\quad g^{}_Z~=~\frac{e}{\cos\theta_W\sin\theta_W}
%~=~\sqrt{(g_1^2+g_2^2)}
\ ,\quad g^{}_W~=~\frac{e}{\sqrt{2}\sin\theta_W}\ .
$$
The electromagnetic current is 

$$
J^{\,\rm em}_\mu~=~\ov {\rm e}^\dagger_i\sigma^{}_\mu \ov {\rm e}^{}_i-{\rm e}^\dagger_i\sigma^{}_\mu {\rm e}^{}_i+\frac{2}{3}{\bf Q}^\dagger_{i1}\sigma^{}_\mu{\bf Q}^{}_{i1}-\frac{1}{3}{\bf Q}^\dagger_{i2}\sigma^{}_\mu{\bf Q}^{}_{i2}-\frac{2}{3}{\ov\bf u}^\dagger_{i}\sigma^{}_\mu\ov{\bf u}^{}_{i}+\frac{1}{3}\ov{\bf d}^\dagger_{i}\sigma^{}_\mu\ov{\bf d}^{}_{i}\ .
$$
The neutral current is 

$$
J^{(Z)}_\mu~=~L_i^\dagger\frac{{\boldsymbol \tau}_3}{2}\sigma^{}_\mu L^{}_i+{\bf Q}^\dagger_{i}\frac{{\boldsymbol \tau}_3}{2}\sigma^{}_\mu{\bf Q}^{}_{i}
-\sin^2_{}\theta^{}_W\,J^{\,\rm em}_\mu\ ,$$
and the charged current is

$$
J^\pm_\mu~=~L_i^\dagger{\boldsymbol \tau}^\pm_{}\sigma^{}_\mu L^{}_i+{\bf Q}^\dagger_{i}{\boldsymbol \tau}^\pm_{}\sigma^{}_\mu{\bf Q}^{}_{i}
\ .$$
At this stage, all fields are massless. There remains to generate masses for the fermions, and rewrite these currents in terms of mass eigenstates.

Fermion masses are generated through Yukawa couplings to the higgs doublet. We can make a spin zero combination out of two left-handed doublets $\psi$ and $\chi$ as

$$
(\psi\chi)~\equiv~i\psi^T\sigma^{}_2\chi~=~\epsilon^{\alpha\beta}_{}\psi^{}_\alpha\chi^{}_\beta\ ,
$$
(if you like indices). We can form three spin zero combinations which are weak doublets. The \SM~Yukawa coupling is therefore,

$$
{\mc L}^{}_{Yuk}~=~y^{({\bf u})}_{ij}({\bf Q}^{}_i\ov{\bf u}^{}_j)\cdot H+\Big(y^{({\bf d})}_{ij}({\bf Q}^{}_i\ov{\bf d}^{}_j)+y^{(\rm e)}_{ij}({L}^{}_i\ov{e}^{}_j)\Big)\cdot H^*_{}
$$
where 

$$
H^*{}=\begin{pmatrix} h_2^{*}\\ -h^{*}_1\end{pmatrix}\ ,$$ and 
the dot stands for $\tau_2$, the $SU(2)$ CG coefficient. 

Not all three Yukawa matrices ${\bf Y}=y_{ij}$ are physical. Without loss of generality, the charged lepton Yukawa coupling can be taken to be diagonal, leaving 

$$
{\mc L}_{Yu}^{(\rm e)}~=~[m_{\rm e}e^{}(\ov {\rm e}\rm e)+ m_\mu^{}(\ov \mu \mu)+m_\tau^{}(\ov \tau \tau)][1+\frac{h}{v}]\ .$$
A similar analysis can be carried out for the down-like quarks, where we set without loss of generality,

$${\bf Y}^{({\bf d})}_{}~=~{\bf V}^{(-1/3)}_{}{\bf D}^{(-1/3)}_{}{\bf W}^{(-1/3)}_{}\ ,$$
where $\bf D$ is diagonal, and $\bf V$ and $\bf W$ are unitary matrices. $\bf W$ is absorbed by redefining  $\ov{\bf d}$, and disappears from the physics. Similarly, we redefine

$${\bf Q}^{'}={\bf Q}{\bf V}^{(-1/3)}_{}\ ,$$
which yields in the electroweak vacuum,

$$
{\mc L}_{Yu}^{({\bf d})}~=~[m_d^{}(\ov {\bf d}{\bf d})+ m_s^{}(\ov {\bf s} {\bf s})+m_b^{}(\ov {\bf b} {\bf b})][1+\frac{h}{v}]\ .$$
So far so good. Now we do the same for the charge $2/3$ quarks and set 

$${\bf Y}^{({\bf u})}_{}~=~{\bf V}^{(2/3)}_{}{\bf D}^{(2/3)}_{}{\bf W}^{(2/3)}_{}\ .$$
The same procedure yields the mass eigenstates

$$
{\mc L}_{Yu}^{({\bf u})}~=~[m_u^{}(\ov {\bf u}{\bf u})+ m_c^{}(\ov {\bf c} {\bf c})+m_t^{}(\ov {\bf t} {\bf t})][1+\frac{h}{v}]\ ,$$
where we have now

$${\bf Q}^{''}={\bf Q}{\bf V}^{(2/3)}_{}\ .$$
The higgs couples to the mass of all fermions as well. 

Unless ${\bf V}^{({\bf u})}={\bf V}^{({\bf d})}$, something has to give. Indeed, the mismatch between these two matrices is physical, the CKM matrix
$$
{\mc U}_{CKM}^{}~=~{\bf V}^{(2/3)}_{}{\bf V}^{(-1/3)\,\dagger}_{}\ .$$ 
It modifies the charged current, which now reads

$$
J^+_\mu~=~{\rm e}^\dagger_i\sigma^{}_\mu \nu^{}_i+{\bf d}^\dagger_{i}({U}^{}_{CKM})^{}_{ij}\sigma^{}_\mu{\bf u}^{}_{j}
\ , $$
 and its conjugate. The neutral currents are not affected. If we take out the diagonal phases from the CKM matrix, we are left with three rotation angles and one $CP$-violating phase (CP violation occurs through complex Yukawa couplings). 

Finally we have the color couplings of the quarks and gluons which preserves parity

$$ 
g^{}_3{\bf G}^{A\mu}_{}{\bf J}^A_\mu\ ;\qquad {\bf J}^A_\mu~=~\frac{1}{2}{\bf Q}^\dagger_{i}{ \lab}^A_{}\sigma^{}_\mu{\bf Q}^{}_{i}+\frac{1}{2}\ov{\bf u}^\dagger_{i}{ \lab}^{*A}_{}\sigma^{}_\mu\ov{\bf u}^{}_{i}+\frac{1}{2}\ov{\bf d}^\dagger_{i}{ \lab}^{*A}_{}\sigma^{}_\mu\ov{\bf d}^{}_{i}
\ .$$
Now that the particles have their requisite masses, it is easy to see that the \SM~ is quark number (baryon number/3) invariant, and, since there are no mixings in the lepton sector, also under the three lepton numbers, one for each family, $L_{\rm e}$, $L_{\mu}$, and $L_{\tau}$. It is convenient to define the total lepton number $L$ as their sum, and two relative lepton numbers, say $L_{\rm e}-L_{\mu}$ and $L_{\mu}-L_{\tau}$, because at the one loop order, neither $B$ nor $L$ are conserved due to an anomaly diagram, only the combination $B-L$. 

This  is the \SM~ in the {\em unitary gauge}, where the Nambu-Goldstone bosons have disappeared. It is very 
physical, but terrible for perturbative calculations. In perturbative gauge theories, we need to add a gauge fixing term to the Action, for instance,

$$\frac{1}{2\alpha}(\partial_\mu A^\mu)^2\ .$$
It is recommended to alter it in spontaneously broken gauge theories. The reason is that the covariant derivative on the scalar field with a vacuum value,
is

$${\mc D}\varphi~=~\partial_\mu\varphi+ieA_\mu \varphi+ievA_\mu\ ,$$
which yields (after integration by parts) a cross term of the form $M{\rm Im}\,\varphi \partial_\mu A^\mu$ in ${\mc L}$. In order to  get rid of this term, 't Hooft introduced the $R_\xi$ gauge with the altered gauge term

$$\frac{1}{2\xi}(\partial_\mu A^\mu+\xi M{\rm Im}\,\varphi)^2\ .$$
In this type of gauge, the Nambu-Goldstone bosons are present and need to be included in the calculation (they have a $\xi$-dependent mass). Physical results should of course be independent of $\xi$. 

\subsection{See How They Run!}
All couplings and parameters of the \SM~run with scale according to the renormalization group equations,

$$
 \frac{d}{dt}\alpha_{}^{-1}~=~\frac{1}{2\pi}(\frac{11}{3}C_{adjt}-\frac{2}{3}C^{}_{Weyl}-\frac{1}{6}C^{}_{Scalar})\ .$$
 
% $$ \frac{d}{dt}\left\{ \begin{array}{rcl}\alpha_1^{-1} \\ \alpha_2^{-1} \\\alpha_3^{-1}
% \left\{ \begin{array}{rcl}-\frac{4}{3}n^{}_{fam}-\frac{1}{10}\ , \\ \frac{22}{3}-\frac{4}{3}n^{}_{fam}-\frac{1}{6} \\11-\frac{4}{3}n^{}_{fam}
%\end{array}\right.$$
 
On their way to the ultraviolet, the three gauge couplings (with three chiral families) evolve as
$$
 \frac{d\alpha_1^{-1}}{dt}~=-\frac{41}{20\pi}\ ,\quad
 \frac{d\alpha_2^{-1}}{dt}~=~\frac{19}{12\pi}\ ,\quad
  \frac{d\alpha_3^{-1}}{dt}~=~\frac{7}{2\pi}
\ .$$
The \SM~ still has the same QED Landau pole, and the evolution of the $SU(2)$ and $SU(3)$ couplings displays asymptotic freedom. 
The Yukawa couplings satisfy equations of the form,

$$\frac{d{\bf Y}^{}_{u,d,e}}{dt}~=~\frac{1}{16\pi^2}{\bf Y}^{}_{u,d,e}{\boldsymbol\beta}^{}_{u,d,e}\ ,$$
which indicate a fixed point at their vanishing, and additional symmetry, which is the  enormous global chiral symmetry $SU(3)^{\bf Q}_L\times SU(3)^{\bf u}_R\times SU(3)^{\bf d}_R\times SU(3)^{L}_L\times SU(3)^{\rm e}_R$.

Keeping only the heaviest third family Yukawa couplings and the strong gauge coupling, these reduce to

$$\frac{dy^{}_\tau}{dt}~=~\frac{3y^{}_\tau}{16\pi^2}y^{2}_t\ ,\quad \frac{dy^{}_b}{dt}~=~\frac{y^{}_b}{16\pi^2}(\frac{3}{2}y^2_t-8g^2_3)\ ,\quad
\frac{dy^{}_t}{dt}~=~\frac{y^{}_t}{16\pi^2}(\frac{9}{2}y^2_t-8g^2_3)\ .$$
Because of the strong coupling, all are also asymptotically free.  

The most interesting and mysterious Higgs self-coupling satisfies an equation of the form

$$
 \frac{d\lambda}{dt}~=~\frac{1}{16\pi^2}\beta^{}_\lambda\ ,$$ 
 where $\beta_\lambda$ does not vanish at $\lambda=0$. It follows that this is point is not {\em natural} in the sense that it has no enhanced symmetry. 
 In the same approximation but with $y_{b,\tau}\ll y_t$, we have
 
$$
 \frac{d\lambda}{dt}~=~\frac{3}{4\pi^2}(\lambda^2_{}+\lambda y^2_t- y^4_t)\ .
 $$
 Since the mass squared of the higgs particle is proportional to $\lambda$, its evolution has clear physical consequences. Let us examine it in two limits:
 
 Qualitatively, for a heavy (compared to $v$) higgs,  $\lambda$ is large, and thus $\beta_\lambda$ is dominated by $\lambda$ and thus overwhelmingly positive, resulting in a Landau pole at energies not much larger than the higgs mass. Quantitatively,  for $M_H\approx 170~ GeV$, perturbation theory is valid up to Planck scale. For heavier values of the Higgs mass, strong coupling is reached at much lower energies. 

On the other hand, if the higgs mass is light compared to $v$, $\beta_\lambda$ is dominated by the {\em negative} $y_t^4$, given that the top quark is so heavy. Hence at some scale $\lambda$ will change sign! The meaning is pretty clear- the potential becomes unbounded from below, the reason why this is the stability bound. It only means that the Ginzburg-Landau potential description of spontaneous breaking is not longer valid. Quantitatively, for $M_h\approx 150~ GeV$, that scale is the Planck scale, so that for $150~GeV\lesssim M_h\lesssim180~GeV$, these limits are at the Planck scale ({\em Planck Chimney}). This is part of the region which the Tevatron detectors have recently ruled out, and LHC is reducing even more, resulting in nail-biting suspense! If the Higgs is found below the Planck Chimney, there must be new physics at a scale {\em above} the electroweak and {\em below} the Planck scales. 
\vskip .3cm
{\em A light Higgs is the Canari of Beyond the \SM~Physics!}
\vskip .3cm
The prime candidate is supersymmetry, where scalar masses are coupled to fermion masses, and share the same naturalness. There $\lambda$ is replaced by the square of gauge and Yukawa couplings.  
 
\subsection{Partial Symmetries of the \SM}
Although the \SM~ has a definite set of symmetries, parts of its Lagrangian may have a larger symmetry. Parity is a prime example of a partial symmetry. QED is parity invariant but it is part of the \SM~which is not. Hence although the bulk of atomic physics is parity invariant, there will be parity violating \SM~effects, which will manifest itself by level splittings. These are small because the interactions are small. It could also be that a tree-level part of the Lagrangian has some symmetry, while the rest does not, but quantum corrections will bring symmetry breaking effects at loop order. 

An example consider tree-level gauge interactions. The charged W-bosons are the only gauge bosons which cause transitions between quarks of different flavors, because of the CKM matrix. Flavor-changing tree level interactions change the electric charge as well, so that there are no tree-level ``neutral current" interactions which change flavors. Of course, virtual particle exchanges will generate interactions where flavor changes but the electric charge does not. These new effects  will appear as irrelevant interactions, and their strength will be finite and calculable. They come under the name of {\em Flavor Changing Neutral Current effects (FCNC)}! The \SM~is not just a pretty tree theory; it is a full-fledge quantum field theory, tested by experiment through its small FCNC loop effects.

Another interesting ``accidental symmetry" is the {\em custodial symmetry}. In the Higgs potential, the four real Higgs fields appear in the combination $\sum_{i=1}^4h_i^2$, which is $SO(4)$ invariant. All students know that $SO(4)$ is $SU(2)\times SU(2)$. The first $SU(2)$ is the gauged weak YM theory, what happens to the second $SU(2)$? It is broken by some Yukawa couplings, namely the lepton Yukawas and part of the quarks'. It is also broken by $g_1$ couplings. Hence it is not broken as long as 

$$g_1^{}~=~0 \, \quad {\bf Y}^{(\rm e)}~=~0\ ,\quad    {\bf Y}^{(\bf u)}~=~{\bf Y}^{(\bf d)}\ .$$
What happens to this symmetry after spontaneous breaking? It is broken down to a vectorial $SO(3)$ subgroup, under which $W^+_\mu,\,W^-_\mu,\, Z^{}_\mu$ form a triplet, resulting in equal strength for charged and neutral current-current interactions at tree level, or

$$\rho~\equiv~\frac{M^2_W}{M^2_Z\cos^2\theta_W}~=~1\ ,\qquad \rho-1~=~0\ .$$
Therefore, the deviation of $\rho$ from this value is expected at loop order, as the custodial symmetry is badly broken by the large $t-b$ mass difference and a little bit by electromagnetism.
 
\subsection{Calculable Interactions }
We can proceed in the same way as in QED to find all possible irrelevant operators, and anticipate their form before any calculations. It is of course complicated by the fact that the Higgs gets a vacuum value, but it turns out that there are only a few combinations of Higgs fields we need to consider with the following  $SU(2)_Y$ properties:

$$
H^\dagger H~\sim~{\bf 1}^{}_0\ ,\quad H^\dagger\vec\tau H~\sim~{\bf 3}^{}_0\ ,\quad (H\, H)~\sim~{\bf 1}^{}_2\ ,\quad H\,H\,H~\sim~{\bf 4}^{}_3\ .$$
and the conjugates for which the hypercharge changes sign. All of these have non-vanishing vacuum values. Invariants which appear in the Lagrangian simply have quantum numbers conjugate to those: we have invariants and covariants. How can we build these? Let $\psi$ be a generic left-handed fermion 
\vskip .2cm
\noindent {\em Dimension-five Invariant Interactions}. $(\psi\psi) HH$. The spin zero two-fermion combination must be a weak triplet, which leaves only one candidate interaction 
$$(L_iL_j)\cdot H\vec\tau H$$ 
symmetrized over the family indices. It satisfies all the gauge symmetries, but remember that lepton number is conserved, so that it will not be generated in perturbation theory! More on this later of course! 
\vskip .2cm
\noindent {\em Dimension-five Covariant Interactions}. We can start by forming  combinations which have the quantum numbers of the Higgs doublet; when coupled with it, they form dimension six interactions. Among the many such combinations, we only note those which do not occur at tree level. For example we have 

$$\ov{\bf s}^\dagger_{}\sigma^{}_\mu\ov{\bf b}\,{\bf\mc  D}^{}_\mu\,H\ ,\qquad \ov{\bf d}^\dagger_{}\sigma^{}_\mu\ov{\bf s}\,{\bf\mc  D}^{}_\mu\,H\ ,
$$
which flavor changing neutral interactions. We also have  magnetic moment interactions

$${\bf Q}^{}_i\sigma^{}_{\mu\nu}{\lab}^A\ov{\bf d}^{}_j\,{\bf G}^{A\,\mu\nu}_{}\ ,\quad {\bf Q}^{}_i\sigma^{}_{\mu\nu}{\tau}^a\ov{\bf d}^{}_j\,{ W}^{a\,\mu\nu}_{}\ ,\quad {\bf Q}^{}_i\sigma^{}_{\mu\nu}\ov{\bf d}^{}_j\,{B}^{\mu\nu}_{}\ ,
$$
some of which change flavor, and the same with ${\bf d}\rightarrow {bf u}$. What makes these interactions interesting is that some of those violate flavor and induce processes like ${\bf b}\rightarrow {\bf s}+\gamma$, etc... .
\vskip .2cm
\noindent {\em Dimension-six Invariant Interactions}. There are many such interactions with three field strengths which are not forbidden by any symmetry such as 

$$\epsilon^{abc}_{}W^a_{\mu\nu}W^b_{\nu\rho}W^c_{\rho\mu}\ ,\qquad \epsilon^{abc}_{}W^a_{\mu\nu}W^b_{\nu\rho}\widetilde W^c_{\rho\mu}\ ,
$$ 
and so-on. We can also make bizarre interactions such as

$$H^\dagger_{}\tau^a_{}HW^a_{\mu\nu}B^{\mu\nu}_{}\ ,$$
which in the vacuum seem to create mixing between $W^3_\mu$ and $B^\mu$! Can you understand this? Interactions of the form $\psi^\dagger {\mc D} {\mc D} {\mc D}\psi$, yield for example,

$$
{\bf d}^\dagger_i\sigma^{\mu}_{}\ov{\bf d}^{}_j{\mc D}^\nu_{}B^{}_{\mu\nu}\ ,$$
and many more; they produce {\em chirality preserving} flavor changing interactions; these are generated by {\em Penguin diagrams}.

Finally, we consider four-fermion interactions. One can form lots of invariants of that form, but many violate baryon number and/or lepton number. We list only a few

$$
{\bf Q}^{}_i{\bf Q}^{}_j\ov{\bf d}^{}_k \ov{\bf u}^{}_l\ ,\quad {\bf Q}^{}_i \ov{\bf u}^{}_j L^{}_k\ov{e}^{}_k\ ,\quad {\bf Q}^{}_i \ov{\bf d}^{}_j {\ov L}^{}_k{e}^{}_l\ .$$
These occur through box diagrams, but unlike QED, they generate flavor-chaging and charge preserving interactions, such as, ${\bf s}\ov{\bf d}\rightarrow  \ov{\bf s}{\bf d}$, which leads to processes that change flavor by two units of flavor; for instance $K-\ov K$, $B-\ov B$ oscillations, etc ... .

Finally, what happens to custodial symmetry? As expected it leaks: computation of the $\rho$ parameter yields

$$
\rho-1~=~\frac{3G^{}_\mu}{8\pi^2\sqrt{2}}\Big( m^2_t+m^2_b-\frac{2m^2_t m^2_b}{m^2_t-m^2_b}\ln\big(\frac{m^2_t}{m^2_b}\big)+M^2_W\ln\big(\frac{M^2_h}{M^2_W}\big)-M^2_Z\ln\big(\frac{M^2_H}{M^2_Z}\big)\Big)\ ,
$$
and you can check that it vanishes in the custodial symmetry limit ($m_t=m_b;\ ,M_w=M_Z)$

As much as I would like to continue discussing the \SM, and showing you on how to calculate its subtlest of effects, my contract with TASI obligates me to move on to discuss some of its extensions. 

%%%%%%%%%%%%%%%%%%%%%%%%%
\section{Massive Neutrinos}
Neutrinos, the poltergeist particles, remain very mysterious. Neutrinos are nightmare particles for experimentalists; they have no electric charge, hardly interact, and move very fast, which makes them most difficult to pin down. It is no wonder that most neutrino experiments and pronouncements, even from world-class scientists, are wrong (at first), but no s(t)igma is attached to their reputations because of the difficulties involved. 

To give a few examples, in $1911$, no less than Hahn, Meitner, and Von Baeyer,  find that $\beta$ radioactivity is like $\alpha$ radioactivity, a two-body process! They found that the produced electron came out at a fixed energy. In 1914, Chadwick and Rutherford expressed doubt, but the war intervened as Chadwick was kept prisoner in Berlin.  It was not until the middle twenties that the issue was resolved: the electrons come out on the average with a fraction of the available energy, eliciting desperate suggestions that energy was not conserved in microscopic processes (Bohr and Slater)! In 1929, Pauli suggests there must exist a spin one half particle {\em inside  the nucleus} that is released in $\beta$ decay. It also solves a recently discovered puzzle with Nitrogen's Raman lines, but is shortly explained by Chadwick's discovery of the neutron. It is Fermi who correctly understands that Pauli's particle, which he calls neutrino,  does not live in nuclei, but is spontaneously created in the decay. Nobody at that time thinks that neutrinos will ever be detected (Today, the detection of the bath of primeval neutrinos predicted by Big Bang cosmology looks equally unlikely). 

In 1956, Cowan and Reines find evidence for the antineutrino produced at the Savannah River nuclear reactor (antineutrinos are produced in $\beta$ decay of matter). Neutrinos are the only elementary particle discovered south of the Mason-Dixon line! At the same time, using Pontecorvo's 1945 suggestion that a neutrino impinging on  cleaning fluid ($CCl_4$) can create an easy-to-detect inert Argon isotope with one month half-life, Ray Davis finds a neutrino (not an anineutrino!) coming out of the same reactor! Pontecorvo, suggests neutrino-antineutrino oscillations (in analogy with Gell-Mann and Pais theory of $K-\ov K$ oscillations), but the effect goes away.  While Ray Davis devotes the rest of his life to the detection of neutrinos generated in the Sun's core, Maurice Goldhaber, in a fiendlishly clever table top experiment, finds  that neutrinos are left-handed, thereby unraveling the origin of parity violation in $\beta$ decay. 

In the intervening years, more features emerged: there is one neutrino species for each charged lepton, and the hypothesis of neutrino masses neatly explains various epochal experiments on solar neutrinos, neutrinos produced by cosmic rays, antineutrino reactors, and today using accelerators to beam neutrinos to distant detectors.  As expected there are experimental anomalies, and a number of people expect surprises of fundamental importance coming from neutrino physics, such as violation of sacred principles, such as CPT and Lorentz invariance! Smart money says that these will go away, but of course you never know! Now for the physics:

By now, you have heard the news: neutrinos of one flavor morph into neutrinos of another flavor, and neutrinos must have different masses, with at least two of them massive. It also means that two global symmetries, $L_e-L_\mu$ and $L_\mu-L_\tau$ are broken. Given the present data, the one remaining global symmetry of the \SM~is $(B-L)$. In addition, cosmology restricts the sum of neutrino masses to be less that $.7$ $eV$! 

So, not only do we need to  incorporate neutrino masses in the \SM, but also explain their tiny values compared to the electroweak natural scale, without affecting the \SM.

What does Lorentz invariance say about neutrino masses? Consider a collection of Weyl Grassmann fields $\psi_{\alpha\,i}$, where $\alpha=1,2$ is the Lorentz spinor index and $i$ is a taxonomic index. Weyl Quadratic combinations split into two types (remember $SU(2)$ group theory: ${\bf 2}\times{\bf 2}={\bf 3}+{\bf 1}$). The first combination

$$
(\overrightarrow{\psi_i\psi_j})=\psi_{\alpha i}\psi_{\beta j}(\tau_2\vec\tau)^{\alpha\beta}$$ 
is $[ij]$ {\em antisymmetric}, and couples to self dual field strengths, such as $(\vec E+i\vec B)$; it describes magnetic and electric dipole moments, and does not interest us at this stage. The second Lorentz invariant combination

$$(\psi_i\psi_j)=\psi_{\alpha i}\psi_{\beta j}\epsilon^{\alpha\beta}_{},$$
describes the $(ij)$ {\em symmetric Majorana mass}. If $\psi$ carry any quantum number, the Majorana mass will carry it twice. 
How do we describe the mass of a charged particle? In  Weyl language, the electron is described by two fields $e$ and $\ov e$, one with charge $(-1)$, the other with charge $(+1)$. The most general bilinear Lorentz invariant is a $(2\times 2)$ matrix

$$\begin{pmatrix} e& \ov {e }\end{pmatrix}
\begin{pmatrix} m_{ee} & m_{e\ov e}\\ m_{\ov {e}e} & m_{\ov{e}\,\ov{e}}\end{pmatrix}\begin{pmatrix} e\\ \ov {e }\end{pmatrix}
$$
But neither $m_{ee}$ nor $m_{\ov ee}$ can describe masses since their carry two units of charge; only the off-diagonal term is a good candidate, since it is electrically neutral. In indices

$$
m_{\ov {e}e}e\,\ov e~=~m_{\ov {e}e}\,e^{}_L\sigma_2\sigma_2e^{*}_R~=-m_{\ov {e}e}\,e^\dagger_Re^{}_L.
$$
In the Lagrangian, the conjugate must be added. This is the familiar {\em Dirac mass}.  As for tango, it takes two spinors to make a Dirac mass, but it only takes one to make a Majorana mass. Unlike the Majorana mass, the Dirac mass does not break fermion (in this case lepton) number symmetry.
This is all you need to know, but we elaborate a little bit.

What kind of mass do neutrinos have, or more to the point, is total lepton number broken? The data does not tell us, although the experimental limits of neutrinoless double $\beta$ decay 

$$
N\rightarrow N'+e^-+e^-\ ,$$
where $N$ and $N'$ are two nuclei, keeps improving, albeit very slowly. 

\SM~neutrinos are Weyl fields with a $SU(2)$ spinor index and hypercharge $-1/2$. Their Majorana mass therefore transforms as a weak triplet
with hypercharge $-1$. As noted in the previous instruction, such a combination can be manufactured by a Higgs bilinear, leading to the dimension-five interaction (suppressing Lorentz indices)

$$
(L_i\tau_2\vec\tau L_j)\,H^T_{}\tau_2\vec\tau H\ .
$$
Since it violates total lepton number by two units: it is not generated in the \SM,  but   
{\em  any extension of the \SM~which violates lepton number necessarily generates Majorana neutrino masses at loop order}. This results in many ways to build \SM~extensions with tiny neutrino masses.   

\begin{itemize}

\item The poor person's extension requires each \SM~neutrino to have a right-handed partner, the same way an electron does. Their resultant Dirac mass does not violate lepton number. In the Pati-Salam extension of the \SM~ (see later instruction), this is exactly what happens and all particles have right-handed partners, assembling in representations of a hitherto unknown (very) weak right-handed $SU(2)$.  This is a fine mechanism, but it offers no explanation for their small value. 

\item A more credible alternative, called the {\em see-saw} mechanism, relies on a large energy scale to generate small neutrino masses. Neutrinos do have right-handed partners, but the right-handed partners have {\em very} large Majorana masses of the order of the Grand Unified scale.This {\em tree-level} suppression  (augmented by radiative corrections coming from the dimension-five operator), engenders neutrino Majorana masses which of course violate total lepton number. 

\item The third class of models relies solely on radiative corrections. All  require new scalar fields endowed with lepton number. Their interactions are rigged to violate lepton number, and therefore unleash at some order of perturbation theory the dimension-five interactions. 

\end{itemize}

\noindent {\bf The See-Saw Mechanism}

\noindent Consider one \SM~neutrino $\nu$ and its right-handed partner $\ov N$. At tree-level, the entries of the resulting $(2\times 2)$ Majorana mass matrix have the following electroweak quantum numbers 

$$
\begin{pmatrix} \Delta I^{}_W=1 & \Delta I^{}_W=\frac{1}{2}\\ \Delta I^{}_W=\frac{1}{2} & \Delta I^{}_W=0\end{pmatrix}
$$
In the absence of a Higgs weak triplet, the $11$ entry is zero (at tree level only); the off diagonal entries are the Dirac masses, which are of the order of say charged lepton and quark masses, that is as most a few hundred $GeV$. The $22$ entry is not protected by any quantum numbers, and it is natural it will be as large as it can be in the theory; for a Grand Unified theory, $10^{14-16}$ $GeV$. In terms of the very small ratio of these entries

$$
\epsilon~=~\frac{\Delta I^{}_W=\frac{1}{2}}{ \Delta I^{}_W=0}~\ll~0,
$$
this matrix has one tiny eigenvalue and one large eigenvalue

$$m^{}_\nu\sim \epsilon^2M\ ,\qquad m^{}_{\ov N}\sim M.
$$
This is the gut of the see-saw mechanism. Neutrino masses are depressed from that of their charged counterparts by the ration of the electroweak to GUT scales. For three neutrinos and their right-handed counterparts, the $(6\times 6)$ symmetric and complex Majorana mass ${\mathcal M}$ is diagonalized by a unitary transformation

$$
{\mc M}~=~{\mc U}^T_{}{\mc D}{\mc U}\ ,\qquad {\mc U}^\dagger_{}~=~{\mc U}^{-1}_{}.
$$
Notice the transpose and not the dagger. Set

$$
{\mc U}~=~\begin{pmatrix} {\mc U}^{}_{11} & \epsilon\, {\mc U}^{}_{12}\\ \epsilon\, {\mc U}^{}_{21} & {\mc U}^{}_{11}\end{pmatrix}\ ,\qquad 
{\mc D}~=~\begin{pmatrix} {\mc D}^{}_{\nu} & 0\\ 0 & {\mc D}^{}_{\ov N}\end{pmatrix}.
$$
Both $ {\mc U}^{}_{11}$ and $ {\mc U}^{}_{22}$ are unitary up to ${\mc O}(\epsilon^2)$ corrections. 

The lepton charged current, written in terms of mass eigenstates becomes,

$$
J^+_\mu~=~{\rm e}^\dagger_{}\sigma^{}_\mu{\mc U}^{}_e({\mc U}^\dagger_{11} \nu^{}_{}+\epsilon{\mc U}^\dagger_{21}\ov N)
%+{\bf d}^\dagger_{i}({\mc U}^{}_{CKM})^{}_{ij}\sigma^{}_\mu{\bf u}^{}_{j}
\ . $$
The low energy physics is described by the almost unitary MNSP lepton mixing matrix

$$
{\mc U}^{}_{MNSP}~=~{\mc U}^{}_e{\mc U}^\dagger_{11}\ .
$$
We can extract its phases by using the same Iwazawa decomposition

$$
{\mc U}^{}_{MNSP}~=~{\mc P}{\mc U}^{'}_{MNSP}{\mc P}^{'}_{}.
$$
Counting parameters, ${\mc U}^{'}_{MNSP}$ contains three rotation angles, and one $CP$ violating phase, as in CKM. The diagonal phase matrix  $\mc P$ can be absorbed in the right-handed charged lepton fields, but the other $\mc P'$ cannot because the neutrino masses are Majorana. It means that in the see-saw, we have three $CP$ violating phases; one is like that in the quark sector, and the other two, linked to the Majorana nature of the masses, can only be detected in lepton number violating processes. 
\vskip .3cm
\noindent {\bf Leptonic Higgs}

\noindent If scalar particles contain lepton number, it must be that they couple to leptons. Assuming only \SM~leptons, there are few possibilities. These Higgs can couple to three Lorentz invariant lepton combinations,  

$$
(L^{}_{(i}\tau^{}_2\vec\tau L^{}_{j)})\cdot\vec T^{}_{2}\ ,\quad (L^{}_{[i}\tau^{}_2 L^{}_{j]})\,S^+_{2}\ ,\quad (\ov e^{}_{(i}\,{\ov e}^{}_{j)})S^{--}_{-4}
$$
resulting in three possible leptonic Higgs. Two of them, the  weak triplet $\vec T=(T^{++},T^{+},T^{0})$, and the  charged weak singlet $S^{--}$ couple symmetrically to the families, while $S^+$ couples antisymmetrically. Since most of these models predict doubly charged scalars, they have been the focus of much experimental scrutiny. 

Explicit lepton number violation in three ways, through cubic self-interations

$$
S^{--}_{}S^+_{}S^+_{}\ ,\quad \vec T\cdot\vec T\,S^{--}_{}\ ,\quad H^T_{}\tau^{}_2\vec\tau H\cdot \vec T,
$$
which violate lepton number by two units. The quartic interactions do not violate lepton number. Finally, we note that in models with two Higgs, lepton number violation can be introduced via 

$$
 H^T_{}\tau^{}_2 H^{'}_{}\,S^+_{}.
 $$
All these models generate Majorana neutrino masses at loop order, of strength dictated by the dimensionfull cubic coupling of the Leptonic Higgs. Further investigate of these models is left to your enjoyment. 

Finally, I have not discussed the possibility of spontaneous breaking of global lepton number. These leave behind a massless Nambu-Goldstone boson called the Majoron. As a  Nambu-Goldstone boson, it couples to the divergence of the (broken) current. In the context of these leptonic Higgs, it is the triplet Majoron, whose existence contradicts the value of the $Z$-boson width. With sufficient cleverness, one can invent singlet Majoron models, but suffice it to say that any new massless particle does violence to physics, and intrudes nefariously in all types of phenomenology. 

  %%%%%%%%%%%%%%
\section{Why Three Families?}
Most  parameters of the  \SM~come from the Yukawa sector which is responsible for generating masses and mixing angles. In the quark sectors, there are six quark masses to be explained, three mixing angles and one $CP$-violating phase. In the original \SM, only the masses of the three charged leptons awaited explanations. With the advent of neutrino oscillations, we now have three neutrino masses, three mixing angles and one (or three) $CP$-violating phases if lepton number is conserved (violated). 

The first flavor mystery is the number of chiral families. Like a bureaucracy, Nature requires fermions in triplicate without any apparent reason. We should note the wild, yet attractive, string speculation which links this number to the topology of the coset manifold required to go from ten to four spacetime dimensions! More conventional ideas associate this number with the dimension of representations of a putative global {\em family group}. If there is such a group, it is necessarily broken, with the absence of flavor-changing neutral currents effects puts stringent limits on its value. 

There are subtelties about masses: they are not constant, and have to be defined at the appropriate scale. For example, even the electron mass, whose value appears in textbooks, is its value at rest. That value does change with the scale at which it is measured (what is its value at the $Z$ pole?), but practically it does not matter since these are very tiny electroweak effects. 

For quarks, the situation is quite different; for one we cannot isolate and ``weigh'' them. The closest we have is the proton and neutron masses. Since each have three quarks, it is tempting to say that each up and down quark weighs a third of a nucleon mass (``constituent quark masses"). On the other hand, we know that QCD contains (by dimensional transmutation) a fundamental scale $\Lambda_{QCD}$, roughly the scale at which its coupling becomes ``strong", commensurate with the proton size and mass, which has nothing to do with the quark mass parameters in the QCD Lagrangian (``current quark masses"). These indeed represent quark 
masses, but in their free limit, {\em  at very short distances}. We can use these values as boundary conditions in their renormalization group equations, and trace them back towards the infrared. We find the top, bottom, and charm quarks with masses above $\Lambda_{QCD}$, with physical values at half the threshold for quark antiquark pair production (I conveniently skip over their dependence on the renormalization scheme). The up, down, and strange quarks have masses below $\Lambda_{QCD}$.  Although we cannot follow each quark mass below the QCD scale, the mass ratios are immune to QCD renormalization effects, and can be compared with experiment. The near equality of proton and neutron masses means that the up and down masses are very small compared to $\Lambda_{QCD}$; they are not equal, as in fact the down quark is twice that of the up quark!  Low energy effective Lagrangian and Lattice Gauge Theory techniques have produced pretty accurate values for these masses. The PDG values are, 

\[
m^{}_{\bf u}~\sim~1.7-3.3~MeV;\quad m^{}_{\bf d}~\sim~4.1-5.8~MeV;\quad m^{}_{\bf s}~\sim~80-130~MeV;
\]

\[
\frac{m_{\bf s}}{m_{\bf d}}~\sim~17-22;\qquad \frac{m_{\bf u}}{m_{\bf d}}~\sim~.35-.60;
\]
The heavier quark have masses (in $\ov{MS}$ scheme)

\[\quad m^{}_{\bf c}~\sim~1.18-1.34~GeV;\quad m^{}_{\bf b}~\sim~4.19^{+.18}_{-.06}~GeV;\quad m^{}_{\bf t}~\sim~172.0\pm.9\pm 1.3~GeV;
\]

\vskip .3cm
\noindent  The $(3\times 3)$ quark mixing matrix (aka the {\bf CKM matrix} is almost equal to the unit matrix, up to corrections of order of the Cabibbo angle $\lambda\sim .22$, and all quark mixing angles are small. It is conveniently described \`a la Wolfenstein, 

\[{\mc U}^{}_{CKM}~=~\begin{pmatrix} 1-\frac{\lambda^2}{2}&\lambda&A\lambda^3(\rho-i\eta)\\-\lambda&1-\frac{\lambda^2}{2}&A\lambda^2\\
A\lambda^3(1-\rho-i\eta)&-A\lambda^2&1
\end{pmatrix},
\] 
where  in terms of four parameters, $\lambda$, $A$, $\rho$ and $\eta$; it is unitary to order $\lambda^3$, and $\eta$ represents $CP$ violation. 

The charged lepton masses are very well determined
 
 \[
\quad m^{}_{e}~\sim~.511~MeV,\quad \quad m^{}_{\mu}~\sim~105.7~MeV, \quad m^{}_{\tau}~\sim~1.77682\pm 0.00016~MeV.\]

With massive neutrinos, mixing in the lepton sector is written in terms of the lepton mixing matrix MNSP (Maki, Nakagawa, Sakata and Pontecorvo). Of its three lepton mixing angles, one is small or zero, and the other two are large. Hence a commonly accepted Wolfenstein-like parametrization does not exist. We write it as a product of three $(2\times 2)$ rotations, ignoring for the moment $CP$ violating phases, 

\[ 
{\mc U}^{}_{MNSP}~=~{\mc R}^{}_{1}(\theta_{23}){\mc R}^{}_{2}(\theta_{13}){\mc R}^{}_{3}(\theta_{12})
\]
where

\[
\theta_{12}~=~34^\circ\pm 1.0^\circ,\quad \theta_{23}~=~45.6^\circ\pm 3.5^\circ \quad \theta_{13}~\le 12^\circ 
\]
(Recent experiments suggest that $\theta_{13}$ is more like $7-8^\circ$). Since neutrino masses are only known through interference, we only know the differences of their mass squared,

\[
\Delta m^2_{12}~ = ~(7.59 \pm 0.21)  \times 10^{-5} eV^2,\qquad  \Delta m^2_{23}  = (0.00243 \pm 0.00013)  \times 10^{-3} eV^2,
\]
which leads to two possible mass orderings (in both cases $m^{}_{\nu^{}_2}\ge m^{}_{\nu^{}_1}$, as deduced from matter effects),

\[
{\rm inverted~ hierarchy}:~m^{}_{\nu^{}_2}\ge m^{}_{\nu^{}_1}\gg m^{}_{\nu^{}_3};\quad ~{\rm normal~ hierarchy}:~m^{}_{\nu^{}_3}\gg m^{}_{\nu^{}_2}\ge m^{}_{\nu^{}_1}
\]
As of this writing, there are hints for a non-zero value of $\theta_{13}$, which would give an extra source of $CP$ violation in the lepton sector. 

\vskip .3cm
\noindent {\bf Theories of Flavor}
Flavor theories are a matter of taste. They are based on numerical relations which could be coincidental or else indicative of deep theory.  Any mixing matrix such as $Y^{-1/3}$ must be diagonalized. Its can be decomposed in terms of its eigenvalues (down quark masses), and mixing angles. Mixing angles and eigenvalues are independent of one another. Yet, as first observed by Gatto, the data suggest a numerical relation between the Cabibbo angle and the square root of the down to strange quark masses ratio,

\[
\tan \theta_{\bf ds}~\sim~\lambda~\sim~\sqrt{\frac{m_{\bf d}}{m_{\bf s}}}.
\] 
which is surprisingly accurate. To see how such a relation can arise, consider the diagonalization of a (real, symmetric) $(2\times 2)$ matrix,

\bean
\begin{pmatrix} a&b\\b&d\end{pmatrix}&=&\begin{pmatrix}\cos\theta&-\sin\theta\\\sin\theta&\cos\theta\end{pmatrix}\begin{pmatrix} m_1&0\\0&m_2\end{pmatrix}\begin{pmatrix}\cos\theta&\sin\theta\\-\sin\theta&\cos\theta\end{pmatrix}\\
&=&\begin{pmatrix} m_1\cos^2\theta-m_2\sin^2\theta&(m_1-m_2)\cos\theta\sin\theta\\(m_1-m_2)\cos\theta\sin\theta&m_1\sin^2\theta-m_2\cos^2\end{pmatrix}
\eean
Inserting the Gatto relation yields

\[
\frac{\sin^2\theta}{\cos^2\theta}~=~\frac{m_1}{m_2},~~\rightarrow~~a~=~0.\]
An obvious interpretation of this {\em texture zero} is family symmetry. The family group forbids this coupling at tree level, but allows it through higher dimension operators. 

Applying the Wolfenstein expansion to quark masses, the data yields a definite hierarchical structure,

\[
\frac{m_{\bf d}}{m_{\bf b}}~=~{\mc O}(\lambda^4),\quad \frac{m_{\bf s}}{m_{\bf b}}~=~{\mc O}(\lambda^2);\qquad 
\frac{m_{\bf u}}{m_{\bf t}}~=~{\mc O}(\lambda^8),\quad \frac{m_{\bf c}}{m_{\bf t}}~=~{\mc O}(\lambda^4).\]
The charged leptons have a hierarchy similar to that of the down-like quarks 

\[
\frac{m_{e}}{m_{\tau}}~\sim~{\mc O}(\lambda^{4-5}),\quad \frac{m_{\mu}}{m_{\tau}}~\sim~{\mc O}(\lambda^2).
\]
There seems to be also hierarchy between some particles of the same family,

\[
\frac{m_{\bf b}}{m_{\bf t}}~=~{\mc O}(\lambda^3),\]
but the quark-lepton ratio

\[
\frac{m_{\bf b}}{m_{\tau}}~\sim~2.3~
\]
is not hierarchical. However, it varies widely with scale, and at the GUT (or later Planck) scale, it is roughly equal to one. This equality can be explained by a simple Higgs structure of $SU(5)$ (see later). 
 
The Gatto relation, which generalizes to the other families, are indicative of some structure. Suppose there is a family group, which punches tree-level holes in the Yukawa matrices. These  texture zeroes will be filled up either by radiative effects or simply as a result of the breaking of the family group, through for instance by a dimension $4+n$  interaction. It will fill the texture zero by $\epsilon^4$, where $\epsilon$ is some small parameter (Froggatt-Nielsen type).   

The quark and charged leptons Yukawa matrices are 

\[
\begin{pmatrix}0&0&0\\
0&0&0\\
0&0&1
\end{pmatrix}
\]
in a tempting  approximation, but we do not know if the heaviest family mixes with the two massless families; if mixing goes as mass ratios, the CKM matrix is the unit matrix, and the third family is stable. 

Is there such an approximation for neutrinos?  The MNSP matrix contains TWO large angles, which implies that at least two massive neutrinos in the same approximation, meaning that either $m_{\nu_3}\gg m_{\nu_2}=m_{\nu_1}$ or $m_{\nu_3}\ll m_{\nu_2}=m_{\nu_1}$. 
 
Let us mention two popular starting matrices with two mixing angles, the ``tribimaximal" matrix, (first introduced by Kaus and Meshkov for quark matrices)  is a ``pretty matrix with an ugly name"\footnote{L. Everett} 

\[
{\mc U}^{}_{MNSP}~=~
\begin{pmatrix}\sqrt{\frac{2}{3}}&-\sqrt{\frac{1}{3}}&0\\
\sqrt{\frac{1}{6}}&\sqrt{\frac{1}{3}}&-\sqrt{\frac{1}{2}}\\
\sqrt{\frac{1}{6}}&\sqrt{\frac{1}{3}}&\sqrt{\frac{1}{2}}
\end{pmatrix}
\]
and a matrix built out of the famous Golden Ratio (prized by architects), written in terms of the Phidian $\phi=(1+\sqrt{5})/2;~\phi^2=1+\phi$,

\[
{\mc U}^{}_{MNSP}~=~
\begin{pmatrix}\sqrt{\frac{\phi}{\sqrt{5}}}&-\sqrt{\frac{1}{\sqrt{5}\phi}}&0\\
\frac{1}{\sqrt{2}}\sqrt{\frac{1}{\sqrt{5}\phi}}&\frac{1}{\sqrt{2}}\sqrt{\frac{\phi}{\sqrt{5}}}&-\frac{1}{\sqrt{2}}\\
\frac{1}{\sqrt{2}}\sqrt{\frac{1}{\sqrt{5}\phi}}&\frac{1}{\sqrt{2}}\sqrt{\frac{\phi}{\sqrt{5}}}&\frac{1}{\sqrt{2}}
\end{pmatrix}
\]
Modern Theories strive to explain these matrices in terms of a family group. If we require the family group to have one, two and  three-dimensional representations, it is natural to seek them among  the finite subgroups of $SU(3)$, $SU(2)$ or $SO(3)$. These have been tabulated by mathematicians a century ago. 

It turns out to be easy to generate these matrices with finite groups.  Subgroups of $SO(3)$, such as the tetrahedral group $A_4$, or the icosahedral group $A_5$, have only three-dimensional representations. In order to explain quark hierarchies, it is necessary to take their double covers (binary extensions) to obtain complex two-dimensional representations (In one particular such model due to KT, all CG coefficients are complex, implying tree-level CP-violation). Finite subgroups of $SU(3)$, which  do not have  two-dimensional representations, are a challenge to explain quark hierarchies. All these models assume specific vacuum structures which generate the initial MNSP matrices. Suffice it to say that they are works in progress. 

It would be pleasing if the family group, like the gauge groups, would turn out to be (in some sense) exceptional. If so, the natural candidate is my pet group $\mc {PSL}_2(7)$ and its subgroups, the projective linear group of $(2\times 2)$ matrices over the Galois field with seven elements. It is a simple $SU(3)$ subgroup with 168 elements, and occupies a central place in mathematics (Klein's quartic curve, Fano Plane, etc...)  

It is unsettling that the mixing patterns of quarks and leptons are quite different even though their quantum number patterns are very similar, as we are about to see in  the last instruction.   

%%%%%%%%%%%%%%%%%%%%%%%
\section{Grand Unification}
QCD's asympotitic freedom, the weakness of interquark dynamics, suggested to Pati and Salam that quarks and leptons are just a gauge group away:  quarks have color, leptons do not. To realize this idea, they generalize color $SU(3)^c$ to hypercolor $SU(4)$. Group theory tells us that

\[
SU(4)~\supset SU(3)^c\times U(1),\qquad {\bf 4}~=~{\bf 3}_{1/3}^c+{\bf 1}^{}_{-1}, \quad {\bf\ov 4}~=~{\bf\ov 3}_{-1/3}^c+{\bf 1}^{}_{1},\]
where the subscript indicates the $U(1)$ value represented by a traceless diagonal matrix. It is ${\mc B}-{\mc L}$, baryon minus total lepton number!

 With hypercolor, each color triplet (left-handed) quark has its color singlet lepton counterpart. Applied to the \SM, it means that, while $\bf\ov d$ is paired with $\ov e$, $\bf Q$ is paired with $L$,  $\bf\ov u$ is paired with a {\em new} lepton we call $\ov N$.

If hypercolor is gauged, new and old gauge bosons live in its fifteen-dimensional adjoint representation. We determine its content via the following group theory

\bean
{\bf 15}+{\bf 1}&=&{\bf 4}\times {\bf \ov 4}\\&=&
[{\bf 3}_{1/3}^c+{\bf 1}^{}_{-1}][{\bf\ov 3}_{-1/3}^c+{\bf 1}^{}_{1}]\\&=&{\bf 3^c\times \ov 3^c}^{}_{(1/3-1/3)}+{\bf 3}^c_{(1/3+1)}+{\bf\ov 3}^c_{(-1/3-1)}+{\bf 1}^{}_{(1-1)}\\
&=&{\bf 8}_0^{c}+{\bf 1}^{c}_0+{\bf 3}^{c}_{4/3}+{\bf\ov 3}^{c}_{-4/3}+{\bf 1}^{c}_0
,
\eean
from which

\[{\bf 15}~=~{\bf 8}_0^{c}+{\bf 1}^{c}_0+{\bf 3}^{c}_{4/3}+{\bf\ov 3}^{c}_{4/3}\]
We recognize the color octet of  gluons, the ${\mc B}-{\mc L}$, and a color triplet of {\em leptoquark} gauge bosons and its conjugate. In the Pati-Salam hypercolor scheme, $SU(4)$ is broken down to color $SU(3)^c$. The leptoquark gauge bosons and that which couples to $\mc B- \mc L$ are very massive. Baryon number is broken, opening the way for proton decay! 

In the \SM, left-handed quarks and leptons form a weak doublet. With the extra lepton $\ov N$, we can imagine that the right-handed quarks and leptons also assemble into doublets of a new extra weak gauged $SU(2)_R$. With the appropriate generalization of the Gell-Mann Nishijima formula

\[
Q~=~I^{}_{3\,L}+I^{}_{3\,R}+(\mc B-\mc L),\]
the \SM~gauge group generalizes to the left-right symmetric structure,

\bean
SU(3)^c\times SU(2)_L\times U(1)_Y&\longrightarrow& SU(3)^c\times SU(2)_L\times SU(2)_R\times U(1)_{(\mc B-\mc L)}\\
&\longrightarrow& SU(4)\times SU(2)_L\times SU(2)_R.
\eean
The rank (number of commuting generators) of this gauge group is five, two more than that of the \SM, and each family now contains sixteen Weyl fermions. It is still characterized by three independent couplings. The \SM~electroweak breaking becomes more complicated, since it has to be generalized to include that of $SU(4)$ and $SU(2)_R$. 

Remarkably, the \SM~fits like a glove in the spinor representation of $SO(10)$ which is complex and sixteen-dimensional (Fritzsch-Minkowski,and Georgi)!  

The Pati-Salam idea of quark-lepton parity can be realized with a rank four gauge group, the (Georgi-Glashow) $SU(5)$. As its name indicates, it is the group that leaves invariant the quadratic form $|z_1|^2+|z_2|^2+|z_3|^2+|z_4|^2+|z_5|^2$, with the group acting on the five complex variables, $z_1,z_2,z_3,z_4,z_5$. Can we find five among the fifteen Weyl fermions of the \SM? We have two five-dimentional combinations,

\[
[(\,{\bf 1}^c\,,\,{\bf 2}\,)^{}_{-1}
+  (\,{\bf \ov 3}^c\,,\,{\bf 1}\,)^{}_{\frac{2}{3}}]
~~~{\rm and}~~~[(\,{\bf 1}^c\,,\,{\bf 2}\,)^{}_{-1}
+  (\,{\bf \ov 3}^c\,,\,{\bf 1}\,)^{}_{-\frac{4}{3}}],
\]
which differ by their total hypercharge. For the first one the hypercharge trace vanishes, which means that it is an element of $SU(5)$.  We do not concern ourselves with the second combination, called {\em split} $SU(5)$, with hypercharge beyond $SU(5)$. 

We identify,

\[
SU(5)~\supset~SU(3)\times SU(2)\times U(1);\qquad {\bf\ov 5}~=~(\,{\bf 1}^c\,,\,{\bf 2}\,)^{}_{-1}
+  (\,{\bf \ov 3}^c\,,\,{\bf 1}\,)^{}_{\frac{2}{3}}.
\]
We still need to account for ten Weyl fermions. Group theory tells us that the antisymmetric product of two quintets is a ten-dimensional representation

\[
({\bf\ov 5}\times {\bf\ov 5})_a~=~{\bf\ov {10}}.
\]
Explicitly,

\[
([(\,{\bf 1}^c\,,\,{\bf 2}\,)^{}_{-1}
+  (\,{\bf \ov 3}^c\,,\,{\bf 1}\,)^{}_{\frac{2}{3}}]\times [(\,{\bf 1}^c\,,\,{\bf 2}\,)^{}_{-1}
+  (\,{\bf \ov 3}^c\,,\,{\bf 1}\,)^{}_{\frac{2}{3}}])_a=(\,{\bf 1}^c\,,\,{\bf 1}\,)^{}_{-2}+(\,{\bf  3}^c\,,\,{\bf 1}\,)^{}_{\frac{4}{3}}+ (\,{\bf \ov 3}^c\,,\,{\bf 2}\,)^{}_{-\frac{1}{3}}.
\]
These describe the left-over fermions, provided that we take the conjugate, that is

\[
{\bf{10}}~=~(\,{\bf 1}^c\,,\,{\bf 1}\,)^{}_{2}+(\,{\bf\ov  3}^c\,,\,{\bf 1}\,)^{}_{-\frac{4}{3}}+ (\,{\bf  3}^c\,,\,{\bf 2}\,)^{}_{\frac{1}{3}}.
\]
Each \SM~chiral family is described by two representations, ${\bf 5}+{\bf\ov{10}}$, which are tucked neatly inside the $SO(10)$ spinor:

\[
SO(10)~\supset~SU(5)\times U(1);\qquad {\bf 16}~=~{\bf\ov 5}+{\bf\ov{10}}+{\bf 1}.
\]
The singlet is of course the $\ov N$ lepton. 

Most features of Grand-Unified theories are best studied in the context of $SU(5)$. To complete its Lagrangian, we need the gauge bosons which are found in the adjoint representation. It is found in the product 

\bean
{\bf 5}\times{\bf\ov 5}&=&{\bf 1}+{\bf 24}\\
&=&[(\,{\bf 1}^c\,,\,{\bf 2}\,)^{}_{1}
+  (\,{\bf  3}^c\,,\,{\bf 1}\,)^{}_{-\frac{2}{3}}]\times [(\,{\bf 1}^c\,,\,{\bf 2}\,)^{}_{-1}
+  (\,{\bf \ov 3}^c\,,\,{\bf 1}\,)^{}_{\frac{2}{3}}]\\
&=&(\,{\bf 1}^c\,,\,{\bf 1}\,)^{}_{0}+\big[(\,{\bf 8}^c\,,\,{\bf 1}\,)^{}_{0}+(\,{\bf 1}^c\,,\,{\bf 3}\,)^{}_{0}+(\,{\bf 1}^c\,,\,{\bf 1}\,)^{}_{0}\big]+(\,{\bf 3}^c\,,\,{\bf 2}\,)^{}_{-\frac{5}{3}}+(\,{\bf \ov 3}^c\,,\,{\bf 2}\,)^{}_{\frac{5}{3}}.
\eean
As we can see from the above, the extra gauge bosons 

\[
{\bf X}_\mu~\sim~ (\,{\bf  3}^c\,,\,{\bf 2}\,)^{}_{-\frac{5}{3}},\qquad {\bf \ov X}_\mu~\sim~ (\,{\bf \ov 3}^c\,,\,{\bf 2}\,)^{}_{\frac{5}{3}},
\]
couple the lepton doublet with the right-handed down quarks

\[
\Big((\,{\bf 1}^c\,,\,{\bf 2}\,)^{}_{1}(\,{\bf \ov 3}^c\,,\,{\bf 1}\,)^{}_{\frac{2}{3}}\Big)
(\,{\bf  3}^c\,,\,{\bf 2}\,)^{}_{-\frac{5}{3}}:~~L^\dagger\sigma_\mu{\bf\ov d}\,{\bf X}_\mu
\]
The adjoint representation is also to be found in 

\[{\bf 10}\times {\bf\ov{10}}~=~{\bf 1}+{\bf 24}+\cdots,
\]  
yielding two other  couplings

\[
\Big((\,{\bf  3}^c\,,\,{\bf 1}\,)^{}_{\frac{4}{3}}(\,{\bf  3}^c\,,\,{\bf 2}\,)^{}_{\frac{1}{3}}+(\,{\bf \ov 3}^c\,,\,{\bf 2}\,)^{}_{-\frac{1}{3}}(\,{\bf 1}^c\,,\,{\bf 1}\,)^{}_{2}\Big)
(\,{\bf  3}^c\,,\,{\bf 2}\,)^{}_{-\frac{5}{3}}:~~\Big({\bf\ov u}^\dagger\sigma^\mu {\bf Q}+{\bf Q}^\dagger \sigma^\mu { \ov e})\Big)\,{\bf X}_\mu
\]
using $({\bf 3}\times {\bf 3})_a={\bf\ov 3}$. 

We see why the $X$ bosons have to be pretty massive: they cause {\em proton decay at tree level} through the chain

\[
\rm Proton~\sim~{\bf u}{\bf d}{\bf u}~~\rightarrow~~ {\bf \ov X}{\bf u}~~\rightarrow~~ e\,{\bf \ov u}{\bf u}~\sim~ e^++\pi^0_{}.\]  
It follows that the amplitude for proton decay is proportional to

\[
\frac{g^2}{M^2_{\bf X}},\]
where $g$ is the $SU(5)$ coupling constant. There is no evidence for proton decay, setting its lifetime at least $10^{31}$ years:  $g$ must be very small, and/or $m_{\bf X}$ be very large. 

Reducing $SU(5)$ down to the \SM, can be achieved by adding a  scalar field, $\Sigma^a_{\,b}$ (I put indices for those of you who like indices $a=1,2,3$ for color, and $4,5$ for weak $SU(2)$), transforming as the adjoint ${\bf 24}$, and arrange its potential so that it acquires a vacuum value.   If we do things properly, the vacuum value $\Sigma=M\,{\rm diagonal}(2,2,2,-3,-3)$ does the job, and the proton lifetime goes as $M^4$.

Grand-Unified extensions of the \SM~entail two vastly different masses, the Higgs mass, and that of the uneaten portion of the $\Sigma$, the Gut-Higgs. In order to avoid proton decay, their masses must be vastly different, 

\[
\frac{M_{Higgs}}{M_{Gut-Higgs}}~\sim 10^{-14}_{}~GeV.
\]
Maintaining this very small ratio in the face of fierce radiative corrections is the famous {\em gauge hierarchy problem}.

The three gauge coupling ``constants" of the \SM~  evolve according to the renormalization group. If $SU(5)$ is to make sense, they need to merge into the $SU(5)$ coupling at some scale.

The values of the fine structure constant and the Weinberg angle  at the electroweak-breaking scale ($\sim$ the $Z$ pole) determine $g_1$ and $g_2$ at that scale, where all three are perturbative. These are such that

\begin{itemize}

\item The $U(1)$ coupling, $g_1$, is very small and increases with energy.

\item The $SU(2)$ coupling, $g_2$, is larger, and decreases slightly with energy.

\item The $SU(3)$ coupling, $g_3$, the largest of the three, decreases sharply with energy. 
  
\end{itemize}
From their qualitative behavior, all three couplings will meet at higher energy, but the question is whether all three meet at one energy or else they meet two at a time. The originally measured value of the Weinberg angle suggested they would meet at one point around $10^{16}$ $GeV$, leading to a definite prediction for the proton lifetime, now ruled out by experiment. The modern value of $\theta_W$ shows that in the absence of any new matter between the electroweak and GUT scale, the couplings narrowly miss one another! Thus in spite of its initial beautiful structure, $SU(5)$, as originally formulated has been ruled out by experiment. 

New matter, the supersymmetric partners of the \SM's particles with masses of the order of $1~TeV$ would put the couplings on the right track. Perhaps the LHC experiments will soon throw light on this subject. 

The $SU(5)$ extension of the \SM~Yukawa sector is designed to generate the quark and charged lepton masses by electroweak breaking.  These are to be found in two places: $
{\bf\ov 5}\cdot{\bf 10}$, which contains the masses of the down-like quarks (charge $-1/3$)and integer charged leptons; ${\bf 10}\cdot {\bf 10}$ with the masses of the up-like quarks (charge $2/3$). 
Group theory says (with indices added)

\[
{\bf\ov 5}\times{\bf 10}~=~{\bf 5}+{\bf 45}: ~~{\bf\ov 5}_{}^a\,{\bf 10}^{}_{bc}~=~\delta^a_{[b}{\bf 5}^{}_{c]}+{\bf 45}^a_{\,bc}
\]

\[
{\bf 10}\times{\bf 10}~=~({\bf \ov 5}+{\bf \ov{50}})_s+{\bf \ov{45}}_a: ~~\,{\bf 10}^{}_{ab}{\bf 10}^{}_{cd}~=~\epsilon^{}_{abcde}{\bf \ov 5}^{e}_{}+\epsilon^{}_{cdef[b}{\bf \ov{45}}^{ef}_{a]}+{\bf \ov{50}}^{}_{abcd}.
\]
Majorana neutrino masses reside in ${\bf\ov 5}\cdot{\bf\ov 5}$,

\[
{\bf\ov 5}\times{\bf\ov 5}~=~{\bf\ov{10}}+{\bf\ov{15}}, \qquad {\bf\ov 5}_{}^a{\bf\ov 5}_{}^b~=~{\bf\ov {10}}_{}^{[ab]}+{\bf\ov {15}}_{}^{(ab)}
\]
Of the five different Higgs fields, only ${\bf 5}, {\bf 45},$ and ${\bf 15}$ can have electrically neutral vacuum values. 

Consider the effect of one Higgs quintet. It contains two fields, the first is a charged scalar color triplet with charge $-1/3$, which was not in the \SM. It must be sufficiently heavy to satisfy the limits on proton decay. Its signature for proton decay is very different from that of gauge-mediated proton decay: since Yukawa couplings couple with mass, it favors decay into the heaviest possible particles, which means strange matter, and a Higgs-mediated proton decay  will produce $K$ meson!

The second field has  the same quantum number as the \SM~Higgs doublet. Its mass is limited by the size of electroweak breaking. This leads to a new problem: how to generate vastly different masses for particles in the same representation\,-\,the {\em Doublet-Triplet  Problem} of Grand-Unified theories. 
    
The Higgs doublet vacuum value does give masses to all charged particles. With one quintet Higgs, $SU(5)$ yields a new relation,  

\[
m^{}_{\bf b}~=~m^{}_{\tau},
\]
which is satisfied, as long as it is taken at the GUT scale; it is a notable success for the theory. Unfortunately, it does not work for the lighter families, as $m_{\bf s}\ne m_{\mu}$, $m_{\bf d}\ne m_{e}$, even at very large energies. There is a way to regain agreement by inserting a selected vacuum value coming from the ${\bf 45}$. 
\vskip .5cm
What does $SO(10)$ bring to this picture? For one, it brings all fermions of one chiral family into its spinor representation. In that sense it is more unified. On the other hand, breaking its gauge symmetries is more complicated, and requires new Higgs scalars. 

The Yukawa couplings are simpler, with 

\[
{\bf 16}\times {\bf 16}~=~{\bf 10}_{sym.}+ {\bf 126}_{sym.}+{\bf 120}_{antisym.}.\]
The decuplet, not to be confused with that of $SU(5)$, is the real vector representation. It contains the $SU(5)$ quintet and its conjugate, ${\bf 10}={\bf 5}+{\bf\ov 5}$. 

One new relation arises, relating the mass of the up-like quarks to the neutrino Dirac masses ($L\ov N$) between the lepton doublets and the new singlet field. However, as $SO(10)$ is broken, its Majorana mass $\ov N\,\ov N$ is not protected, and is expected to be  around that scale.  As a result,  neutrino masses are more complicated, mixing a Dirac mass of electroweak magnitude with Majorana mass of order GUT  scale. This is the see-saw mechanism, which predicts that neutrino masses are suppressed from that of their charged lepton partners by the ratio of the electroweak to GUT scale.

There is only one embedding of $SU(5)$ inside $SO(10)$, but there are two ways to identify the \SM~hypercharge. In the original formulation,  hypercharge is inside $SU(5)$. In {\em Split} $SU(5)$, hypercharge is not, but still in $SO(10)$.

At the end of this instruction, I would be remiss not to mention $E_6$ the next largest unification group. It is, like Nature, exceptional, and makes the bridge with superstring theories. Its fundamental representation is the complex {\bf 27}, which contains the $SO(10)$ $\bf 16$, $\bf 10$ and one  extra singlet.  

\section{Acknowledgements}
It was a real pleasure to lecture at TASI 2011. My thanks to the TASI students for their many insightful questions, as well as the lecture organizers, Professors K. Matchev and T. Tait. The kind hospitality and efficacy of Professors T. DeGrand and K. T. Mahanthappa, created the right atmosphere for intellectual pursuit and interactions.   
 
 \section{References}
Rather than referencing the original literature, I list below a list of books and monographs which I have found useful, without any attempt at completeness.
\vskip .5cm

\noindent{ \bf Quantum Field Theory}
\vskip .3cm
\noindent{\em Modern Field Theory: A Concise Introduction} by T. Banks (Cambridge University Press, 2008)

\noindent{\em Quantum Field Theory} by C. Itzykson and J.-B. Zuber (McGraw-Hill 1980)

\noindent{\em Quantum Field Theory}, by F. Mandl and G. Shaw (Wiley \& Sons, 1984)

\noindent{\em An Introduction to Quantum Field Theory} by M. Peskin and D. V. Schroeder (Westview Press 1995)

\noindent{\em Field Theory: a Modern Primer} by P. Ramond (Frontiers in Physics, 2nd ed., 1988)

\noindent{\em Quantum Field Theory}, by M. A. Srednicki (Cambridge University Press, 2006)

\noindent{\em The Quantum Field Theory of Fields, Vols 1 \& 2}, by S. Weinberg (Cambridge University Press 1995)

\noindent{\em Quantum Field Theory in a Nutshell}, by A. Zee (Princeton University Press 2003)

\vskip 1cm

\noindent {\bf Standard Model and Beyond}
\vskip .3cm
\noindent{\em Gauge Theories in Particle Physics}, by I.J.R. Atchinson and A.J.G. Hey (Taylor \& Francis 2003)

\noindent{\em Physics of Massive Neutrinos} by F. Boehm and P. Vogl (Cambridge University Press 2nd ed 1992)

\noindent{\em Dynamics of the Standard Model} by J. Donoghue, E. Golowich, and  B. Holstein (Cambridge University Press 1992)

\noindent{\em The Physics of Massive Neutrinos}, by B. Kayser (World Scientific 1989)

\noindent{\em The Standard Model and Beyond}, by P. Langacker (Taylor \& Francis 2010)

\noindent{\em Gauge Field Theories} by Stefan Pokorski (Cambridge University Press 2nd ed 2000)

\noindent{\em Journeys Beyond the Standard Model}, by P. Ramond (Perseus Books, Cambridge 1999)

\noindent{\em Grand Unified Theories}, by G. G. Ross (Frontiers in Physics 1983)
\vskip .3cm

\noindent {\bf Group Theory}
\vskip .3cm

\noindent{\em Semi-Simple Algebras and Their Representations} by R. N. Cahn (Frontiers in Physics 1984)

\noindent{\em Lie Algebras in Particle Physics} by H. Georgi (Frontiers in Physics 1982)

\noindent{\em Theory and Application of Finite Groups} by G.A. Miller, H. F. Blichtfeldt and L. E. Dickson, (John Wiley \& Sons, NY, 1916)

\noindent{\em Group Theory} P. Ramond (Cambridge University Press 2010)

%P. Langacker, Phys. Repts, 72(1981) 185 

%%%%%%%%%%
\end{document}